\documentclass[aps, prd, twocolumn, superscriptaddress, preprintnumbers, nofootinbib, longbibliography]{revtex4-1}

\usepackage{amsmath}	
\usepackage{amsthm}		
\usepackage{amssymb}	
\usepackage{datetime}
\usepackage{graphicx}
\usepackage{color}
\usepackage{verbatim}
\usepackage{bm}

\usepackage[colorlinks=true, citecolor=midblue, linkcolor=midblue, urlcolor=midblue]{hyperref}

\usepackage{setspace}

\usepackage{tikz,xcolor}
\definecolor{lime}{HTML}{A6CE39}
\DeclareRobustCommand{\orcidicon}{
	\begin{tikzpicture}
	\draw[lime, fill=lime] (0,0) 
	circle [radius=0.16] 
	node[white] {{\fontfamily{qag}\selectfont \tiny ID}};
	\draw[white, fill=white] (-0.0625,0.095) 
	circle [radius=0.007];
	\end{tikzpicture}
	\hspace{-2mm}
}
\foreach \x in {A, ..., Z}{
	\expandafter\xdef\csname orcid\x\endcsname{\noexpand\href{https://orcid.org/\csname orcidauthor\x\endcsname}{\noexpand\orcidicon}}
}

\settimeformat{ampmtime}

\definecolor{grey}{rgb}{0.4,0.4,0.4}
\definecolor{dullmagenta}{rgb}{0.4,0,0.4}
\definecolor{darkblue}{rgb}{0,0,0.4}
\definecolor{midblue}{rgb}{0,0,0.5}
\definecolor{midred}{rgb}{0.5,0,0}
\definecolor{orange}{rgb}{1,0.5,0}
\definecolor{lightbrown}{rgb}{0.75,0.5,0.25}
\definecolor{tan}{cmyk}{0.14,0.42,0.56,0}
\definecolor{djunglegreen}{cmyk}{0.99,0,0.52,0}
\definecolor{lightgreen}{rgb}{0,1,0}
\definecolor{olivegreen}{cmyk}{0.64,0,0.95,0.40}
\definecolor{midgreen}{rgb}{0.0,0.675,0.0}
\definecolor{darkgreen}{rgb}{0,0.5,0}

\newcommand{\Tr}{\ensuremath{\mathrm{Tr}}}

\newcommand{\rchi}{\raisebox{\depth}{$\chi$}}

\settimeformat{ampmtime}

\usepackage[font=small,labelfont=bf,justification=raggedright]{caption}
\usepackage{subcaption}

\usepackage[notrig]{physics}

\newcommand{\FirstAffiliation}{\affiliation{
	Arnold Sommerfeld Center,
	Ludwig-Maximilians-Universit{\"a}t,
	Theresienstra{\ss}e 37,
	80333 M{\"u}nchen,
	Germany}}
 
\newcommand{\SecondAffiliation}{\affiliation{
	Max-Planck-Institut f{\"u}r Physik,
	F{\"o}hringer Ring 6,
	80805 M{\"u}nchen,
	Germany}}

 \newcommand{\ThirdAffiliation}{\affiliation{
        Tsung-Dao Lee Institute and School of Physics and Astronomy,
        Shanghai Jiao Tong University,
        Shanghai, China
	}}

\newcommand{\FourthAffiliation}{\affiliation{
 School of Physics and Astronomy, Shanghai Jiao Tong University, Shanghai, China}}

\date{\formatdate{\day}{\month}{\year}, \currenttime}

\begin{document}

\title{
Confinement Slingshot and Gravitational Waves
}

\author{Maximilian Bachmaier\orcidB{}}
\email{maximilian.bachmaier@physik.uni-muenchen.de}
\FirstAffiliation
\SecondAffiliation

\author{Gia Dvali}
\FirstAffiliation
\SecondAffiliation

\author{Juan Sebasti\'an Valbuena-Berm\'udez\orcidA{}}
\email{juanv@mpp.mpg.de}
\FirstAffiliation
\SecondAffiliation

\author{Michael Zantedeschi\orcidC{}}
\email{zantedeschim@sjtu.edu.cn}
\ThirdAffiliation
\FourthAffiliation

\date{\small\today} 

\begin{abstract} 

\noindent In this paper, we introduce and numerically simulate a quantum field theoretic phenomenon called the gauge ``slingshot" effect and study its production of gravitational waves. The effect occurs when a source, such as a magnetic monopole or a quark, crosses the boundary between the Coulomb and confining phases. The corresponding gauge field of the source, either electric or magnetic, gets confined into a flux tube stretching in the form of a string (cosmic or a QCD type) that attaches the source to the domain wall separating the two phases. The string tension accelerates the source towards the wall as sort of a slingshot. The slingshot phenomenon is also exhibited by various sources of other co-dimensionality, such as cosmic strings confined by domain walls or vortices confined by $Z_2$ strings.   
Apart from the field-theoretic value, the slingshot effect has important cosmological implications, as it provides a distinct source for gravitational waves. The effect is expected to be generic in various extensions of the standard model such as grand unification.

\end{abstract}

\maketitle
    \section{Introduction}
\label{sec:introduction}
Understanding the transition between the confining and deconfining regimes of gauge theories remains one of the most fundamental challenges in physics. An enlightening input in this direction can be provided by the study of systems in which the different phases of a gauge theory can
coexist in a controllable manner. 
The early example is provided by the construction given in \cite{Dvali:1996xe} in which a domain wall (or a vacuum layer) supports a deconfined (Coulomb) phase of a gauge theory that at the same time exhibits the confining behavior in the bulk of space. Such coexistence of phases has some important implications.

In particular, the layer of the deconfined phase localizes a massless $U(1)$ gauge field in the Coulomb regime. 
One effect of this localization is that the charges (e.g., quarks), placed in a bulk of the confining vacuum, become attached to the deconfining boundary (wall) by the QCD flux tubes. The string stretches from the quark towards the boundary and opens up there.  
The flux carried by the string spreads within the deconfined layer in the form of the $U(1)$ Coulomb field. 
The system thereby realizes a field-theoretic analog to a $D$-brane. 

The dual setups, in which the analogous effect is exhibited by the magnetic charges have been constructed in \cite{Dvali:1997sa, Dvali:2002fi, Dvali:2015onv}. In these cases, it is the magnetic flux that is confined in flux tubes (cosmic strings) in one of the vacuum domains. The same flux, in the neighboring domain, gets deconfined and spreads in the form of a Coulomb-magnetic field of a magnetic monopole.
 
In a system with coexisting phases, the interesting question is, what happens when charges (either electric or magnetic) cross the boundary separating the two phases? 
 
In the present paper, we shall study this behavior.
We first consider the case of magnetic charges.    
For this purpose, we construct a prototype $SU(2)$ gauge theory which admits two types of vacua. The vacuum in which $SU(2)$ is Higgsed down to a $U(1)$ subgroup and the one in which the $U(1)$ is further Higgsed. 
The first vacuum supports free 't Hooft-Polyakov monopoles that are magnetically charged under $U(1)$. These monopoles are in the Coulomb magnetic phase.  
 
In the second vacuum, the monopoles are confined, i.e., the monopoles and antimonopoles are connected by magnetic flux tubes. These magnetic flux tubes represent Nielsen-Olesen strings~\cite{Nielsen:1973cs} of the $U(1)$ gauge theory (analogous to Abrikosov flux tubes~\cite{ABRIKOSOV1957199} in superconductors). 
 
A domain wall separates the two vacua. 
In this system, we study a scattering process in which a monopole crosses from the magnetic-Coulomb to the magnetic-confining phase. 
Due to the conservation of the magnetic charge, the magnetic flux follows the monopole into the confining phase. However, the confinement makes the flux trapped in a string. 
 
The monopole thus becomes attached to the boundary wall by the string. The string opens on the wall, releasing the entire flux into the Coulomb vacuum.  
For an observer placed in the $U(1)$ Coulomb vacuum, the end-point of the flux tube carries the entire magnetic charge of the 't Hooft-Polyakov monopole. 
In this way, the monopole that crosses from the Coulomb into a confining domain leaves its ``image" on the boundary. The image is represented by the throat of the same flux tube via which the monopole is attached to the boundary from the opposite side.    
     
One important dynamical question is what happens when the energy in the collision process 
is much larger as compared to the rest mass of the monopole?
A possible outcome one may consider is that the string breaks up by creating monopole-antimonopole pairs. 
Then, instead of penetrating deeply into the confining domain and stretching a long string, the energy of the collision is released in the form of many monopole-antimonopole pairs connected by short strings, which will soon annihilate into waves. In this case, one could say that effectively the magnetic charge never enters the confining domain. 
 
This (naive) intuition is supported by the study of the annihilation of monopoles connected by a string \cite{Dvali:2022vwh}. 
As shown there, after coming on top of each other, the pair does not  oscillate even once. Instead, it decays into the waves of Higgs and gauge fields.   
This effect was explained by entropic arguments: the entropy of a monopole-antimonopole pair is much lower than the entropy of waves. Once the monopole and antimonopole come on top of each other, the system loses the memory of its pre-history of the magnetic dipole. 
After this point, it simply evolves into the highest entropy state, which is given by waves, as opposed to monopoles connected by a long string. In the language of amplitudes, this can be understood as insufficient entropy for compensating a strongly suppressed process of production of highly coherent states~\cite{Dvali:2020wqi}. 

This outcome is characteristic of the phenomenon 
of defect ``erasure", originally discussed in~\cite{Dvali:1997sa} for the interaction 
of monopoles and domain walls. In~\cite{Dvali:2022vwh} was argued that the same effect must hold for heavy quark-antiquark pairs connected by the QCD strings.  
   
As we shall show, in the present case, the situation is very different. 
The reason is the existence of the net un-erased magnetic charge. 
That is, in contrast with the cases of monopole-antimonopole \cite{Dvali:2022vwh} and monopole-wall \cite{Dvali:1997sa}. The net magnetic charge is always point-like. Due to this, the system is aware of its magnetic pre-history all the time. 
Unlike the erasing antimonopole or an erasing domain wall, 
in the present case, the magnetic charge neither cancels nor spreads. Correspondingly, the string never breaks apart. 
     
Thus the outcome is a formation of a monopole attached to a boundary by a long string. The string stretches and absorbs the initial kinetic energy of the monopole, gradually slowing it down. If the wall is static, after reaching a certain maximal size, the string will start shrinking accelerating the monopole towards the boundary. After reaching the boundary, the monopole will be shot back into the Coulomb vacuum. We shall refer to this phenomenon as the ``slingshot" effect.

One of the important implications of the slingshot effect is 
the novel source of production of gravitational waves. 
The monopole slingshot effect is expected to be rather generic in the early cosmology of grand unified theories. It is thereby 
important to understand the imprints of this effect in the gravitational 
wave spectrum. 

Due to this, we study the corresponding gravitational wave signal in detail in our parameters space range. In particular, it is found that the energy spectrum and the beaming angle of emission are analogous to the case of a monopole-antimonopole connected by a string in the confined phase, complying with the fact that most of the signal is due to the acceleration of the monopole by the slingshot. In particular, the energy spectrum is found to scale as the inverse frequency $\omega^{-1}$, which agrees with studies of a confined monopole-antimonopole pair in the point-like approximation~\cite{Martin:1996cp} and in the fully-fledged field theoretical case~\cite{Dvali:2022vwh}. Moreover, we also observe that the slingshot gravitational radiation is emitted in a beaming angle $\theta$, measured from the acceleration axis of the domain wall, scaling approximately as $\omega^{-1/2}$.

The same type of gravitational wave signal is expected in the dual slingshot case of the ``electric" confinement. 
In this case, the role of a monopole is played by a heavy quark that crosses over from the Coulomb to a confining domain. Similarly to the monopole stretching a cosmic string, the quark entering the confining domain stretches a QCD flux tube. 
For explicit analysis
we construct a model using the earlier setup discussed in~\cite{Dvali:2002fi, Dvali:2007nm}  for the study of the gauge field localization mechanism of~\cite{Dvali:1996xe}.
In this setup, the two vacua represent confining and deconfining phases of $SU(2)$ QCD. 
We assume that the quarks are heavier than the corresponding QCD scale. 
    
The QCD flux tubes connect these quarks in the confining domain. The flux tubes can be exponentially long without the danger of breaking apart.  
In the deconfining domain, $SU(2)$ is Higgsed down to $U(1)$, and the same quarks can propagate freely and interact via the Coulomb $U(1)$ field. 
The two phases are separated by a domain wall.
The massless photon is ``locked" in the $U(1)$ domain by the gauge field localization mechanism of \cite{Dvali:1996xe}. 
   
We then consider a scattering process in which a heavy quark goes across the wall from the $U(1)$ Coulomb to the $SU(2)$-confining phase.     
Transporting the intuition from the monopole case of the dual theory, we shall argue that the outcome is similar: the system exhibits a slingshot effect. Namely, the quark stretches along the QCD string which connects it to the wall.  
Despite the sufficient energy in the collision, the string does not break up into a multiplicity of mesons and glueballs. The physical reason, as we shall argue, is similar to the monopole case and has to do with the existence of the net $U(1)$ charge measured by the Coulomb observer.  

Just like the magnetic slingshot effect, its electric dual can be 
relevant for cosmology in various extensions of the standard model,
including grand unification, since the coexistence of 
phases is rather generic. The gravitational wave signal from the electric slingshot is rather similar to its magnetic dual.

Finally, the slingshot effect generalizes to defects of other co-dimensions. In particular, it can be exhibited by cosmic strings. When the string crosses over into a phase in which it becomes a boundary of the domain wall, it stretches the wall. The essence of the effect is captured by an effective $2+1$-dimensional model. The model in which $Z_2$ vortices can be confined was constructed earlier in~\cite{Dvali:1991ka}. 
We extend this model by allowing the coexistence of two phases: the free phase with exact $Z_2$ symmetry, as well as, the confining phase in which $Z_2$ is spontaneously broken. When the vortex crosses over from the free into the confining phase, it results in a slingshot effect. 

The main findings of this paper are summarized in a companion letter~\cite{slinglett}. 
    \section{The Model}
\label{sec:the-model}

We shall now construct a simple prototype model that possesses a domain wall separating the two vacua in which the magnetic field is in Coulomb and confining phases respectively.
Such examples were constructed earlier in~\cite{Dvali:2002fi} as setups for realizing 
a dual (magnetic) version of the gauge field localization mechanism of~\cite{Dvali:1996xe}. 
Correspondingly, in the construction of~\cite{Dvali:2002fi} there exist 
Higgs and Coulomb phases of a $U(1)$ gauge theory are separated by a domain wall. 
Within the $U(1)$ Higgs domain, the magnetic flux is trapped in the tubes (cosmic strings). 
A string can terminate perpendicularly to the wall and open up on the other side in the form of 
the sources of a magnetic Coulomb field. In order to include the magnetic monopoles on both sides of the wall, we embed the $U(1)$ as a subgroup of an $SU(2)$ gauge symmetry.    

The model that we will analyze is an $SU(2)$ gauge theory with two scalar fields. The first field $\phi$ transforms under the adjoint representation, while the second field $\psi$ is in the fundamental representation. The Lagrangian of the theory is
\begin{align}\label{eq:model1}
    \mathcal{L}=&\Tr \left((D_\mu \phi)^\dagger\, (D^\mu \phi)\right)+(D_\mu \psi)^\dagger (D^\mu \psi)\nonumber\\
    &-\frac{1}{2}\Tr \left(G^{\mu\nu}G_{\mu\nu}\right)-U(\phi,\psi)\,,
\end{align}
with the potential
\begin{align}\label{eq:Potential}
    U(\phi,\psi)=&\lambda_\phi \left(\Tr (\phi^\dagger \phi)-\frac{v_\phi^2}{2}\right)^2\nonumber\\
    &+\lambda_\psi \left(\psi^\dagger \psi- v_\psi^2 \right)^2\ \psi^\dagger \psi+\beta \psi^\dagger \phi \psi\,.
\end{align}
The field strength tensor and the covariant derivatives are given in the conventional form
\begin{align}
    G_{\mu\nu}&=\partial_\mu W_\nu - \partial_\nu W_\mu -ig [W_\mu,W_\nu]\,,\\
    D_\mu \phi&=\partial_\mu \phi  -ig [W_\mu,\phi]\,,\\
    D_\mu \psi&=\partial_\mu \psi -ig W_\mu \psi\,.
\end{align}
We can write the gauge field and the adjoint scalar field as $W_\mu =W_\mu^a T^a$ and $\phi=\phi^a T^a$ respectively, where the $SU(2)$ generators are normalized by $\Tr (T^a T^b)=\frac{1}{2}\delta^{ab}$.

In this study, we consider a symmetry-breaking hierarchy characterized by several distinct stages. Initially, the $SU(2)$ symmetry undergoes a Higgs mechanism through the scalar field $\phi$, resulting in the reduction of symmetry to $U(1)$. Subsequently, the $U(1)$ symmetry is Higgsed further down by $\psi$. Schematically, 
the breaking pattern is
\begin{equation}
SU(2)\rightarrow U(1)\rightarrow 1\,.
\end{equation}
During the first breaking process, two of the gauge bosons acquire a mass $m_{v_\phi}=g v_\phi$, while one gauge boson, which we will refer to as the photon, remains massless. The corresponding Higgs boson manifests a mass $m_{h_\phi}=\sqrt{2\lambda_\phi}v_\phi$. Following the second symmetry breaking, all gauge bosons, including the photon, acquire an additional contribution to their mass denoted by $m_{v_\psi}=g v_\psi/\sqrt{2}$. Additionally, the Higgs boson acquires a mass $m_{h_\psi}=2\sqrt{\lambda_\psi} v_\psi^2$ in this subsequent stage.
We note here that although the potential~\eqref{eq:Potential} is non-renormalizable, it does not concern our analysis since such a potential can be obtained from a renormalizable theory by the introduction of an additional gauge singlet field, as it was previously discussed in~\cite{Dvali:2002fi}. Further examples can be found in the same paper.
The classical field equations of this theory are
\begin{align}
    &(D_\mu G^{\mu\nu})^a=j^{a,\nu}_\phi+j^{a,\nu}_\psi\,,\label{eq:field-equation-W}\\
    &(D_\mu D^\mu \phi)^a+\pdv{V(\phi,\psi)}{\phi^a}=0\,,\label{eq:field-equation-phi}\\
    &D_\mu D^\mu \psi+\pdv{V(\phi,\psi)}{\psi^\dagger}=0\,,\label{eq:field-equation-psi}
\end{align}
where the currents are $j^{a,\nu}_\phi=g \varepsilon^{abc}(D^\nu \phi)^b\phi^c$ and $j^{a,\nu}_\psi=i g \psi^\dagger T^a (D^\nu \psi)+h.c.$.

For $\psi=0$, the $SU(2)$ symmetry is Higgsed down to $U(1)$. Consequently, the theory encompasses a magnetic monopole solution characterized by the 't Hooft-Polyakov magnetic monopole ansatz \cite{tHooft:1974kcl, Polyakov:1974ek}
\begin{align}
    W^a_i &= \varepsilon_{aij}\frac{r^j}{r^2}\frac{1}{g} \left(1-K(r)\right),\nonumber\\
    W^a_t &=0\,,\nonumber\\
    \phi^a&=\frac{r^a}{r^2}\frac{1}{g}H(r)\,, \label{eq:tHooft-Polyakov-ansatz}
\end{align}
where $K(r)$ and $H(r)$ are profile functions that depend on the parameters of the theory.

The $SU(2)$ magnetic field can be defined by
\begin{align}
    B^a_k=-\frac{1}{2}\varepsilon_{kij}G^a_{ij}\,.
\end{align}
In order to obtain the $U(1)$ magnetic field, we can project out the component that is parallel to $\phi$.
This yields
\begin{align}
\label{eq:definition-magnetic-field}
    B_k^{U(1)}=\frac{\phi^a}{\sqrt{\phi^b \phi^b}} B^a_k \,.
\end{align}
With this definition, the magnetic field of the 't Hooft-Polyakov magnetic monopole
in the limit of large $r$ is given by,
\begin{align}
    B_k^{U(1)}\rightarrow \frac{1}{g}\frac{r^k}{r^3} \,.
\end{align}

\begin{figure}
  \includegraphics[width=\linewidth]{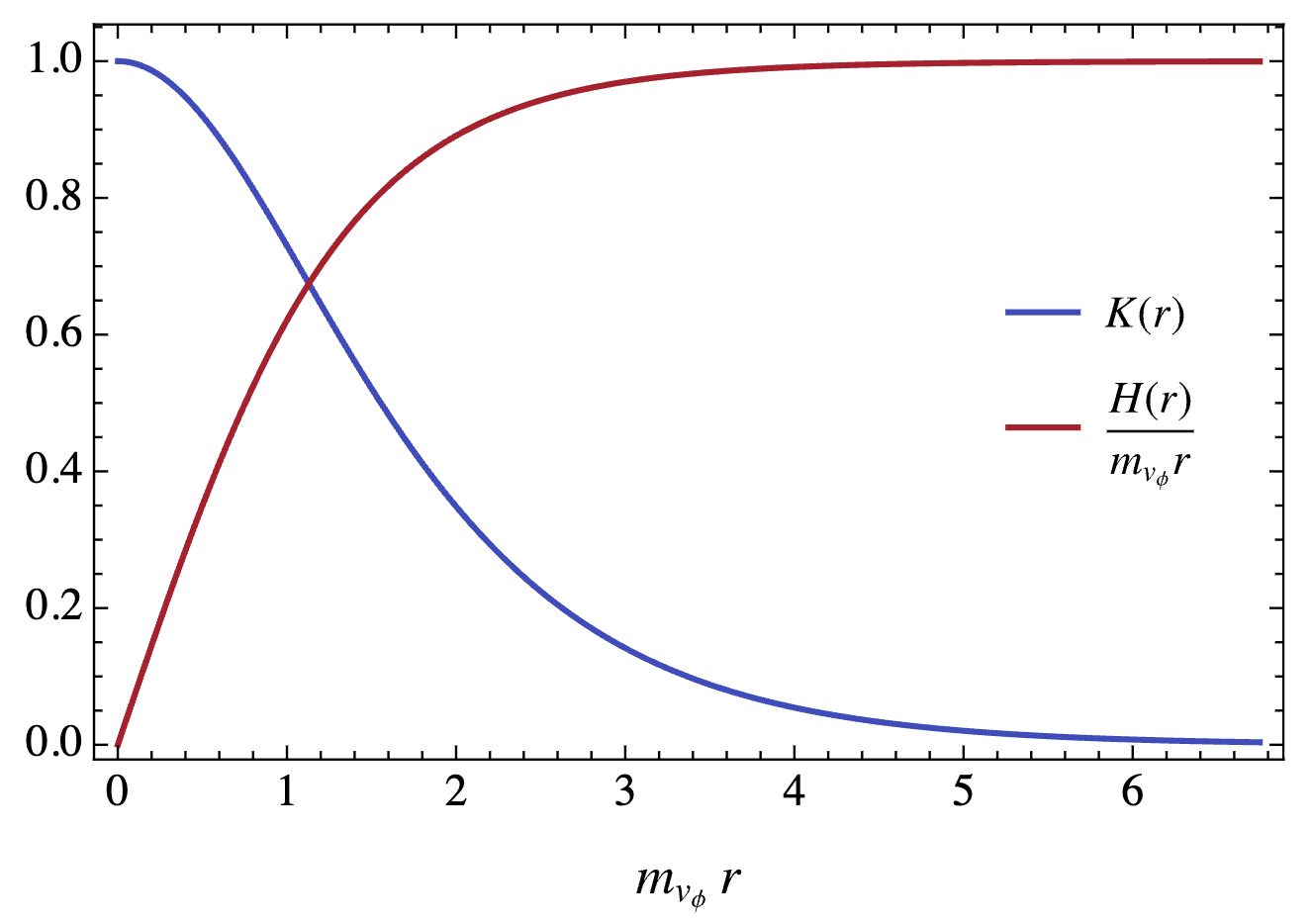}
  \caption{The profile function for a 't Hooft-Polyakov magnetic monopole for $m_{h_\phi}/m_{v_\phi}=1$.}
  \label{fig:profile-functions}
\end{figure}

By substituting the ansatz \eqref{eq:tHooft-Polyakov-ansatz} into the field equations \eqref{eq:field-equation-W} and \eqref{eq:field-equation-phi}, these equations can be simplified to
\begin{align}
    K''=&\frac{1}{r^2}(K^3-K+H^2 K+J^2 K) \,,\\
    H''=&\frac{2}{r^2}HK^2+\frac{1}{2}m^2_{h_\phi}H \left(\frac{H^2}{m^2_{v_\phi} r^2}-1\right)\,.
\end{align}
Note that we are still considering the case with $\psi=0$.
The profile functions can be determined analytically in the BPS limit $m_{h_\phi}\rightarrow 0$~\cite{Bogomolny:1975de, Prasad:1975kr}. For other parameter choices, we employ numerical relaxation techniques. In order to initiate the iteration procedure, we utilize the profile functions obtained in the BPS limit as a starting point.
The resulting profile functions are visualized in Fig.~\ref{fig:profile-functions}.

Let us now shift our focus to the discussion regarding the $\psi$ field.
For the moment let us fix the $SU(2)$ direction to be $\psi=(\xi,0)^T$.
For $\beta =0$ the potential part corresponding to the $\xi$ field exhibits two distinct vacua, the $U(1)$ Coulomb phase at $\xi^\dagger \xi=0$ and the  $U(1)$-Higgsed phase at $\xi^\dagger \xi =v_\psi^2$. The reason behind this terminology will be further explained later. Since these two vacua are disconnected, the model allows a domain wall solution interpolating between them. By using the Bogomolny equation~\cite{Bogomolny:1975de} the solutions can be found to be\footnote{See also~\cite{Gani:2014gxa, Dvali:2022rgx}.}
\begin{align}
    \xi_{(\pm v_\psi,0)}(z)&=\frac{\pm v_\psi}{\sqrt{1+e^{m_{h_\psi}z}}}\,,\\
    \xi_{(0,\pm v_\psi)}(z)&=\frac{\pm v_\psi}{\sqrt{1+e^{-m_{h_\psi}z}}}\,.
\end{align}
Therefore, the two phases, the $U(1)$ invariant and the $U(1)$-Higgsed phase can coexist and are separated by these domain walls.

When $\beta\neq 0$, the degeneracy of the two vacua is broken. The potential difference between the $U(1)$ Coulomb vacuum and the $U(1)$-Higgsed vacuum eliminates the possibility of a static domain wall. This potential difference generates a pressure difference between the two sides of the domain wall, causing it to accelerate toward the phase with higher potential energy. To achieve higher relative collision velocities for our numerical analysis of the interaction between a magnetic monopole and this type of domain wall, we exploit this acceleration. 
Of course, the splitting of energies between different vacua, and thereby the amount of the pressure difference acting on the wall, can be controlled by the parameters of the Lagrangian. In particular, the vacua can easily be kept to be exactly degenerate in energy, resulting in the possibility of static domain walls. 

As a final comment on the spectrum of the theory, we note that once the $U(1)$ is Higgsed down by $\psi$, the free monopole solutions no longer exist. Instead, the monopoles 
get connected to antimonopoles by the cosmic strings.  These strings 
represent Nielsen-Olesen magnetic flux tubes~\cite{Nielsen:1973cs} that carry 
the $U(1)$ magnetic field lines sourced by the monopoles. Since $U(1)$ is embedded in 
$SU(2)$, the strings are not topologically-stable and can break by quantum nucleation of monopole-antimonopole pairs~\cite{Vilenkin:1984ib}. The process is exponentially suppressed by the ratio of the two symmetry-breaking scales. As a result even for a mild hierarchy of scales, an unperturbed segment of a long string is practically stable against such a decay. In particular, this will be the case in our 
analysis.
    \section{Initial Configuration}
\label{sec:initial-configuration}

One generic phenomenon experienced by the monopoles in the confinement regime is the 
annihilation of a monopole-antimonopole pair connected by a string.  
During this process, the monopole and antimonopole are pulled together by the string and subsequently annihilate.
In the approximation of point-like monopoles connected by a thin string, the system is allowed to perform several oscillations. Since in this approximation the structures are not resolved, the monopoles
are permitted to pass through each other and stretch a long string multiple times~\cite{Martin:1996cp}. 
However, the fully resolved analysis shows that this is not the case~\cite{Dvali:2022vwh}.
In the regime of finite and comparable thicknesses of strings and monopoles,
the system decays after the first collision. In~\cite{Dvali:2022vwh} this is explained by the 
loss of coherence~\cite{Dvali:1997sa} during the collision and the entropy suppression 
characteristic for the creation of low entropy solitons in high energy collision processes~\cite{Dvali:2020wqi}.

In the present case, we wish to investigate another type of scattering process involving the 
confined monopole. However, instead of being in the confinement regime from the very beginning, 
initially, the monopole starts in the magnetic Coulomb phase and only later enters the confinement domain with a relativistic velocity.  

Thus, we aim to determine the initial configuration for a specific scenario: a magnetic monopole positioned within the $U(1)$ Coulomb phase, while elsewhere, a domain wall separates the Coulomb phase and the Higgsed phase. We want to analyze in a numerical simulation what happens when the monopole collides with the domain wall. 

In the phase where the $U(1)$ symmetry is Higgsed, the photon, that is massless in the $U(1)$ Coulomb phase, receives a mass. Notice that the magnetic charge is still fully conserved. However, 
in the $U(1)$ Higgs domain the flux can only exist in the form of flux tubes. 
This is energetically costly.  Thereby, the lowest energy configuration with 
a single monopole placed in the $U(1)$ Coulomb domain is the one  
in which the entire flux is spread within the same domain. 
Upon reaching the wall, the magnetic flux lines are repelled and 
spread parallel to the wall.

To include this effect in the initial configuration, we made use of the monopole-antimonopole ansatz with a maximal twist~\cite{Saurabh:2017ryg}. If we take only the monopole side of this ansatz, the magnetic field lines resemble the right behavior. The general ansatz for $\hat{\phi}^a$, where $\hat{\phi}^a=\phi^a/\sqrt{\phi^b \phi^b}$, for a monopole-antimonopole configuration is~\cite{Saurabh:2017ryg}
\begin{align}
    \hat{\phi}_1=&\left(\sin \bar{\theta} \cos \theta-\sin \theta \cos \bar{\theta} \cos \alpha\right)\cos \left(\varphi-\alpha/2\right)\nonumber \\
    &+ \sin \theta \sin \alpha \sin \left(\varphi-\alpha/2\right),\nonumber\\
    \hat{\phi}_2=&\left(\sin \bar{\theta} \cos \theta-\sin \theta \cos \bar{\theta} \cos \alpha\right)\sin\left(\varphi-\alpha/2\right)\nonumber \\
    &- \sin \theta \sin \alpha \sin\left(\varphi-\alpha/2\right),\nonumber\\
    \hat{\phi}_3=&-\cos \theta \cos \bar{\theta} -\sin\theta \sin \bar{\theta}\cos \alpha\,. \label{eq:monopole-antimonopole-ansatz}
\end{align}
The angle $\alpha$ represents the relative twist between the monopole and the antimonopole.
In our simulations, we took $\alpha=\pi$ to obtain a configuration for which the magnetic field presents the right behavior.

We will take the monopoles to be located on the $z$-axis at $z_{\rm M}$ and $z_{\rm \bar{M}}$. Thus, $\varphi$ is the azimuthal angle around the $z$-axis. $\theta$ and $\bar{\theta}$ correspond to the angles between the $z$-axis and the position vectors originating from the monopole and antimonopole, respectively.

The  
ansatz that Saurabh and Vachaspati considered \cite{Saurabh:2017ryg} is then given by
\begin{align}
    \phi^a=&\frac{1}{g}\frac{H(r_{\rm M})}{r_{\rm M}}\frac{H(r_{\rm \bar{M}})}{r_{\rm \bar{M}}}\hat{\phi}^a,\\
    W^a_\mu =&-\frac{1}{g} (1-K(r_{\rm M}))(1-K(r_{\rm \bar{M}}))\varepsilon_{abc}\hat{\phi}^b \partial_\mu \hat{\phi}^c \,.
\end{align}

In order to Lorentz boost this configuration we can replace $z-z_{\rm M}$ and $z-z_{\rm \bar{M}}$ by $\gamma_{\rm M} (z-u_{\rm M} t -z_{\rm M})$ and $\gamma_{\rm M} (z-u_{\rm \bar{M}} t -z_{\rm \bar{M}})$ respectively. For $t=0$, we obtain the initial field values. The values for the time derivatives can be determined numerically by using the field configuration at $t=0$ and $t=\dd t$. 
We conducted the numerical simulations in the Lorenz gauge. 

\begin{figure}
    \includegraphics[width=\linewidth]{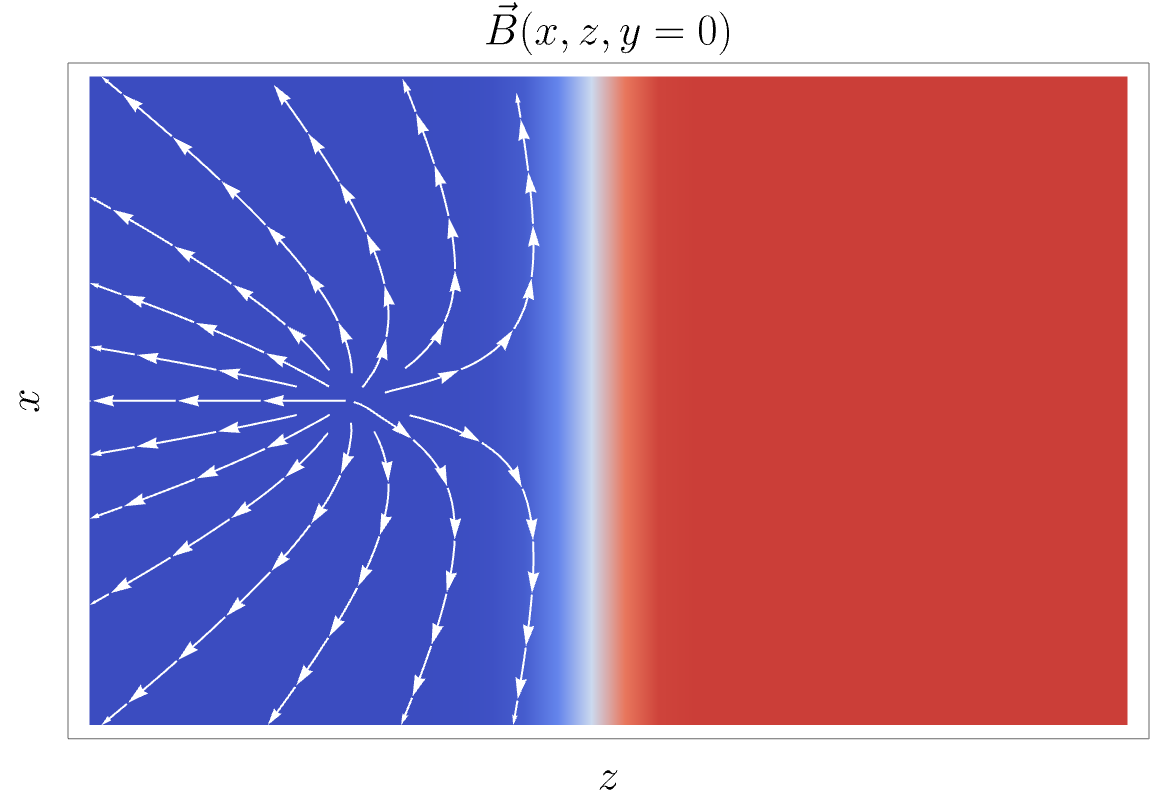}
    \includegraphics[width=\linewidth]{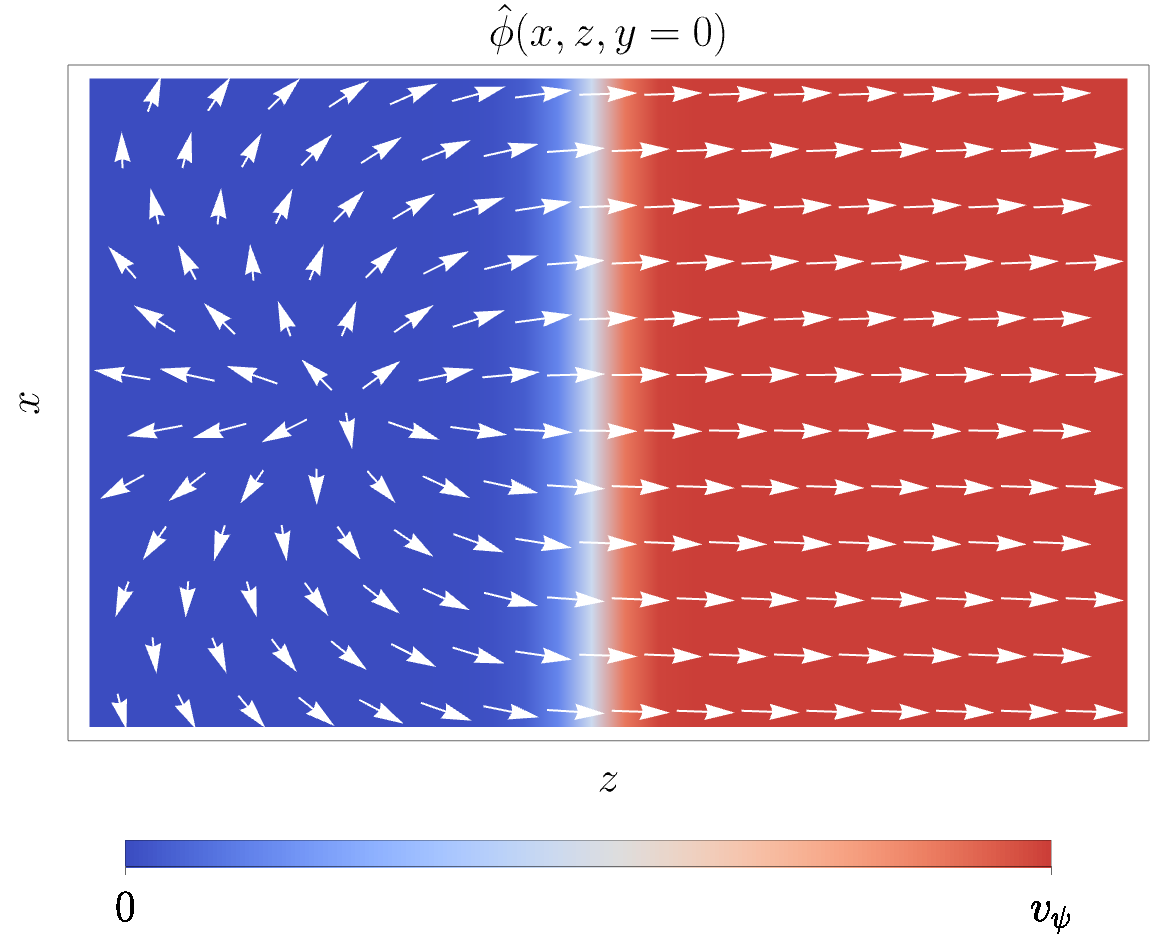}
    \caption{A sketch of the magnetic field (top) and the scalar field vector $(\phi^3, -\phi^2)^T$ (bottom) for the initial configuration. The color in the background represents $\abs{\psi}$ ranging from $\abs{\psi}=0$ (blue) to $\abs{\psi}=v_\psi$ (red).}
    \label{fig:initial-configuration}
\end{figure}

In our configuration, we will incorporate a domain wall positioned at the center between the monopole and the antimonopole. The domain wall is located at $z=0$ and the monopole is on the $z<0$ side. To remove the antimonopole from our setup, we modified our ansatz by
\begin{align}
    \phi^a(x,y,z>0) &\rightarrow \phi^a (x,y,z=0)\,,\\
    W^a_\mu (x,y,z>0) &\rightarrow W^a_\mu (x,y,z=0) \frac{1}{1+e^{\gamma_{\rm D} m_{v_\psi} z}}\,,\label{eq:W-suppression}
\end{align}
where $\gamma_\text{D}$ is the Lorentz factor of the domain wall.
Note that the suppression factor in equation \eqref{eq:W-suppression} is an approximation that is in accordance with the wall profile.

The ansatz for the $\psi$ field that includes the domain wall solution needs to minimize the potential.
In order to achieve this, we seek to extremize the interaction term, 
\begin{align}\label{eq:interaction}
    \mathcal{L}_{int}=-\beta \psi^\dagger \phi \psi\,,
\end{align}
by aligning $\psi^\dagger T^a \psi$ parallel to $\phi^a$.
Therefore, the ansatz for $\psi$ can be written as
\begin{align}
    \psi^1=&-\frac{\xi_{(0,+v_\psi)}}{\sqrt{2}}\frac{(\hat{\phi}^1-i \hat{\phi}^2)}{\sqrt{1+\hat{\phi}^3}}\,,\nonumber\\
    \psi^2=&\frac{\xi_{(0,+v_\psi)}}{\sqrt{2}}\sqrt{1+\hat{\phi}^3}\,.\label{eq:ansatz-psi-field}
\end{align}
The Lorentz boosted configuration of the domain wall can be determined in a similar manner to that of the magnetic monopole. Specifically, we replace the variable $z$ with $\gamma_{\rm D}(z-u_{\rm D} t)$, where $\gamma_{\rm D}$ represents the Lorentz factor associated with the domain wall.

In Fig.~\ref{fig:initial-configuration}, the scalar fields and the magnetic field are illustrated.
Since the ansatz we are employing is an approximation, we incorporated it into a numerical relaxation procedure, as outlined in \cite{Saurabh:2017ryg}, to investigate the response of the field to the static field equations. Notably, we observed that the deviations between the configuration before and after the relaxation remained small, thus affirming its suitability for our intended purpose.
    \section{Numerical Implementation}
\label{sec:numerical-implementation}
The numerical simulations were performed using the Python programming language, leveraging the Numba package \cite{Numba}. Numba facilitates the translation of Python code into efficient machine code, enabling faster computations. Additionally, it offers a straightforward approach to parallelizing the code, effectively utilizing the capabilities of multi-core processors.

In order to improve the computation time, we took benefit of the axial symmetry of the configuration, as we have previously done in the context of magnetic monopole erasure~\cite{Bachmaier:2023zmq}. The approach involved utilizing only three lattice points in the $y$-direction, sufficient for numerically calculating the second-order derivative appearing in the field equations. At each time iteration step, we solved the field equations in the $y=0$ plane and used the axial symmetry to determine the field values in the two neighboring planes. This method, first employed in configurations of this nature, was introduced in~\cite{Pogosian:1999zi}.

From \eqref{eq:monopole-antimonopole-ansatz} we can find the axial symmetry of the $\phi^a$ field of a monopole-antimonopole system for an arbitrary twist. This is given by
\begin{align}
    \phi^1=&f_1 x+f_2 y\,,\nonumber\\
    \phi^2=&f_1 y-f_2 x\,,\nonumber\\
    \phi^3=&f_3\label{eq:axial-symmetry-phi}\,,
\end{align}
where the functions $f_i$ depend only on the time $t$, the radius around the $z$-axis, and the $z$-coordinate. To find an ansatz for the gauge fields we inserted \eqref{eq:axial-symmetry-phi} into $D_\mu \phi=0$. This gives us
\begin{align}
    &W^1_x=xy f_4 +y^2 f_5 +f_6 && W^1_y=y^2 f_4 - f_5 xy-f_7\nonumber\\
    &W^2_x=-x^2 f_4 -f_5 xy+f_7 && W^2_y = -xy f_4 +x^2 f_5 +f_6\nonumber\\
    &W^3_x=x f_8 + y f_9        && W^3_y=y f_8 -x f_9\nonumber\\[0.2cm]
    &W^1_z=f_{10} x+f_{11} y    && W^1_t = f_{12}x+f_{13}y\nonumber\\
    &W^2_z=-f_{11}x+f_{10}y     && W^2_t =-f_{13}x+f_{12}y\nonumber\\
    &W^3_z=0                    && W^3_t=0\,.
\end{align}
From equation \eqref{eq:ansatz-psi-field} we can find the axial symmetric ansatz for the $\psi$ field
\begin{align}
    &\psi^1=f_{14}(x-i y)+f_{15}(y+i x)\,,\nonumber\\
    &\psi^2=f_{16}\,.
\end{align}

Notice that this axial symmetric ansatz presented here can be also used in the analysis of head-on collisions between a monopole and an antimonopole like in the situations described in~\cite{Vachaspati:2015ahr, Dvali:2022vwh}. 

We employed the second iterative Crank-Nicolson method, as described in~\cite{Teukolsky:1999rm}, to simulate the time evolution. We applied the axial symmetry method described above every time we solved the field equation in the $y=0$ plane. We used absorbing boundary conditions for $\phi^a$ and $W_\mu$. For $\psi$ we chose Dirichlet boundaries in $z$-direction and periodic boundaries in $x$-direction. Notice that for the twist $\alpha=\pi$, the imaginary component of $\psi^1$ is anti-symmetric in $x$-direction.

The theory \eqref{eq:model1} contains six 
independent parameters which can be given in terms of: $g$, the masses  $m_{v_\phi}$, $m_{h_\phi}$, $m_{v_\psi}$,  $m_{h_\psi}$, and $\beta$. 
The first three parameters were set to $g=1$ and $m_{v_\phi}/m_{h_\phi}=1$. 
We varied the latter three parameters in the intervals $m_{v_\psi} \in [0.1 \,m_{v_\phi},0.7\, m_{v_\phi}]$, $m_{h_\psi} \in [0.1\, m_{v_\phi}, 1.0\, m_{v_\phi}]$, and $\beta \in [0.001 \,m_{v_\phi}, 0.1\, m_{v_\phi}]$. In the results section we will focus especially on the case with $m_{v_\psi}=0.15\, m_{v_\phi}$, $m_{h_\psi}=0.6 \,m_{v_\phi}$ and $\beta = 0.01\, m_{v_\phi}$.

Besides the aforementioned parameters, we have the flexibility to select the initial velocities of the magnetic monopole ($u_{\rm M}$) and the domain wall ($u_{\rm D}$), as well as the distance between them. In the potential \eqref{eq:Potential}, the interaction term between $\phi$ and $\psi$ causes the domain wall to experience acceleration. Consequently, achieving a collision between the monopole and the domain wall does not necessitate a Lorentz boost. Nevertheless, we varied the initial velocities in the interval $u_{\rm M}, u_{\rm D} \in [0,0.98]$ (in units of $c=1$).
Below we will specifically focus on the scenario where the initial velocities are set to $u_{\rm M}=0.8$ and $u_{\rm D}=0.8$ in opposite directions.
The domain wall was located at $z=0$ and the magnetic monopole at $z=z_{\rm M}=-40\, m_{v_\phi}$.

For the numerical simulations, we used a lattice of the size $[-60\,m_{v_\phi}^{-1},60\,m_{v_\phi}^{-1}]$ and $[-180\,m_{v_\phi}^{-1},60\, m_{v_\phi}^{-1}]$  in $x$- and $z$-direction respectively. The lattice spacing was set to $0.25\,m_{v_\phi}^{-1}$ and the time step we chose to be $0.1\, m_{v_\phi}^{-1}$. The time interval under investigation was $[0, 180\, m_{v_\phi}^{-1}]$.
    \section{Results}
\label{sec:results}
During the time evolution of the initial setup outlined in the previous section, we can observe that as the magnetic monopole approaches the domain wall, a significant amount of magnetic energy density accumulates along the wall. This phenomenon arises due to the presence of a mass for the photon on the right-hand side of the domain wall. As a consequence, the penetration of the photon, which carries the magnetic energy, into the Higgs vacuum is exponentially suppressed. 

Notice, however, that the magnetic field is repelled from the $U(1)$ Higgs domain but not screened~\cite{Dvali:2002fi}.  This is analogous to the Meissner effect in superconductors. 
Some of us already discussed the dual case, in which the electric field is repelled while the magnetic field is screened by a confining layer~\cite{Bachmaier:2023zmq}.
The penetration is possible only in the form of a flux tube, which is costly in energy.
This repulsion leads to the concentration of energy density along the wall, resulting in the observed phenomenon. 

Upon collision with the wall, the monopole transitions to the right-hand side and stretches a string, as this is the only way in which the monopole can enter the $U(1)$ Higgs region. 
The end of the string opens up on the $U(1)$ Coulomb side of the wall,
where the flux can spread out. Since the magnetic charge is conserved, the integrated flux exactly matches 
the magnetic charge of the monopole. Correspondingly, an observer located in the Coulomb 
vacuum will effectively measure the same magnetic charge carried by the string ``throat" 
as the one taken by the original monopole.

\begin{figure*}
    \includegraphics[trim=30 0 60 5,clip,width=\textwidth]{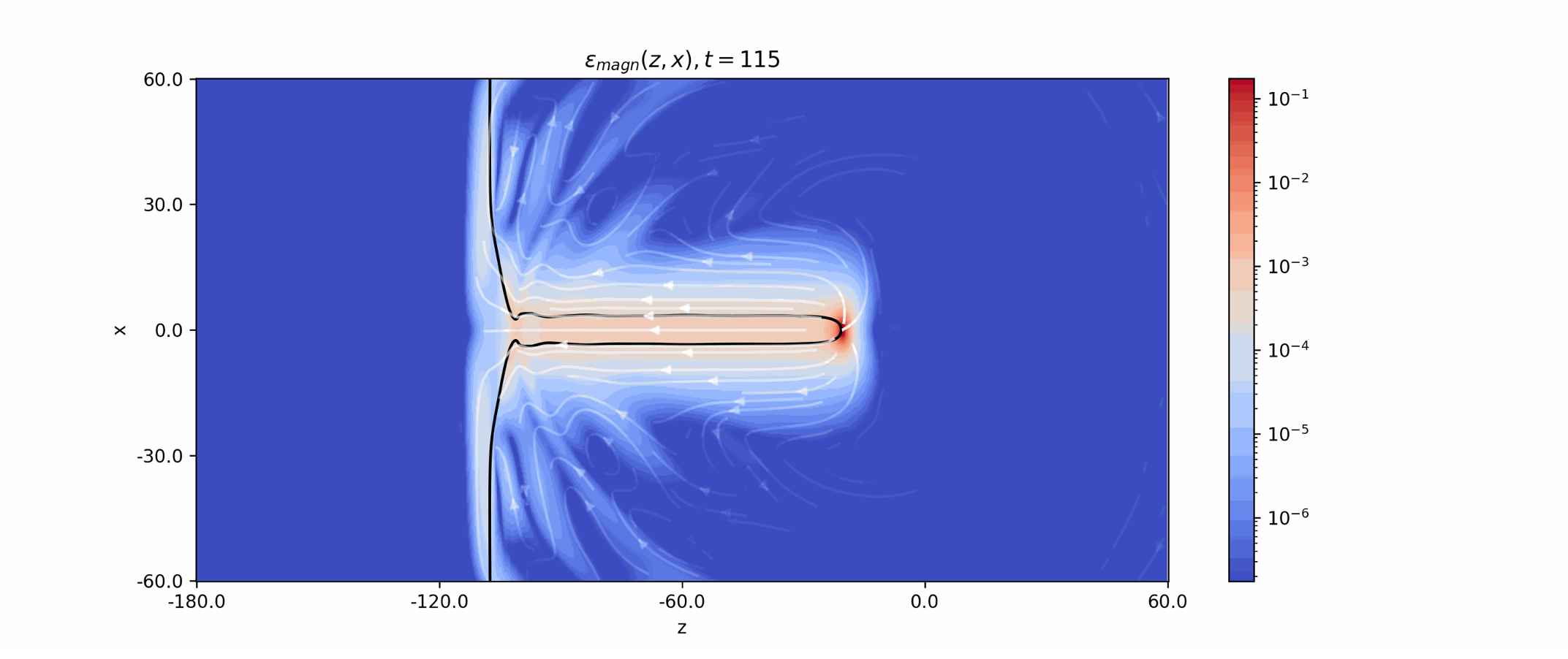}
    \begin{flushleft}
    \caption{
    The illustration depicts the magnetic energy density and magnetic field at time $t=115 \,m_{v_\phi}^{-1}$ in the $y=0$ plane for the specific case described in the numerical implementation section. The length values are provided in units of $m_{v_\phi}^{-1}$, while the energy density values are given in units of $m_{v_\phi}^4/g^2$. The black line represents the contour corresponding to $\abs{\psi}=0.1 \,m_{v_\phi}$, serving to illustrate the presence of the domain wall. We observe that the magnetic monopole has formed a string, connecting it to the domain wall. Both the magnetic field lines and the magnetic energy density indicate the presence of a localized magnetic flux within the string. In the Coulomb phase, the magnetic field vectors point radially away from the point where the string attaches to the domain wall, representing the magnetic field of a virtual monopole located at that particular position.}
    \label{fig:magnetic-energy-density}
    \end{flushleft}
\end{figure*}

This phenomenon can be seen in the magnetic energy density and the behavior of the magnetic field as illustrated in Fig.~\ref{fig:magnetic-energy-density}.
In addition to this figure, the full-time evolution can be found in the video in the ancillary files or at the following link:\\
\url{https://youtu.be/IPJAPjo3nSc?si}

In the time evolution, we can also see that the monopole decelerates during the string stretching. At a certain point, the string approaches its maximum length, and the entire configuration, with the monopole connected to the domain wall via the string, moves collectively at the same velocity. Here it is important to note that it is not exactly the same velocity but approaches the same speed asymptotically. The reason for this is that the domain wall as well as the monopole have a proper constant acceleration,  
\begin{align}\label{eq:accelerations}
a_{\rm DW} &\sim \frac{\delta}{\sigma_{\rm DW}}\sim \beta \frac{m_{v_\phi}}{g m_{h_\psi}}\,,\\
a_{\rm M} &\sim \frac{\mu_{\rm string}}{M_{\rm M}}\sim  \frac{m_{v_\psi}^2}{m_{v_\phi}}\,,
\end{align}
where $\delta$ is the potential energy difference between the $U(1)$-symmetric and the $U(1)$-Higgsed vacua, $\sigma_{\rm DW} \sim m_{h_\psi}v_\psi^2$ is the domain wall tension, and $\mu_{\rm string}\sim m_{v_\psi}^2/g^2$ is the string tension.
Of course, the accelerations in \eqref{eq:accelerations} may differ. In some cases, it is even possible that the monopole is expelled from the Higgsed phase and re-enters it as soon as the domain wall catches up with it again. It is worth noting that the interaction term present in the potential equation \eqref{eq:Potential} plays a crucial role in this behavior. As previously mentioned, this term introduces the vacuum energy difference between the Coulomb and Higgs phases. Consequently, there is a constant acceleration of the domain wall. This acceleration is essential for preventing the monopole from re-entering the Coulomb phase since the tension of the string pulls it outward. Without the domain wall's acceleration, the monopole would be drawn back into the Coulomb phase by the string slingshot effect.

Another notable observation regarding the magnetic energy density is the emission of radiation during the interactions. When the monopole collides with the wall, a significant amount of energy is invested in creating the string, resulting in an extreme deceleration. This process generates electromagnetic radiation in the form of a shock wave. We can also see that this radiation is capable of penetrating into the Higgs phase, demonstrating its ability to traverse regions with broken $U(1)$ symmetry. 

In the parameter regime under consideration, the formation of a string was observed to be nearly ubiquitous. However, when $m_{v_\psi}$ and $m_{h_\psi}$ are sufficiently large, the energy gap at the domain wall becomes too large for the string to form. As a result, the monopole remains localized on the wall and moves together with it. Additionally, the thickness of the string is dependent on the specific parameters of the theory. These parameters and the initial velocities of the monopole and domain wall also determine the maximum length of the string. When these objects possess higher velocities, there is increased availability of energy, allowing the string to extend to greater lengths. As a general estimate, assuming the point-like limit for the monopole solution, the thin string, and the thin wall limit, the maximal penetration is  
\begin{equation}\label{eq:penetration}
\ell_{\rm max} \sim \gamma_c \frac{M_{\rm M}}{\mu_{\rm string}}\,,
\end{equation} 
where $\gamma_c$ is the relative gamma factor between the wall and the monopole at the moment of the collision.\\

The natural question that arises is what is the fate of the extended string. Energetically, it is theoretically possible to form monopole-antimonopole pairs connected by strings after the collision. 
However, despite considering various parameters in our classical simulation, we have not observed this phenomenon. In our simulations, the magnetic monopole consistently stretches the string and maintains its connection to the domain wall, as long as there are no external influences present. Yet, if one perturbed the string, the only way we found so far to disconnect the string from the domain wall is by introducing an additional antimonopole.

In other simulations, we examined a specific configuration where a monopole and an antimonopole, separated by a sufficiently large distance, enter the $U(1)$-Higgsed phase successively along the $z$-axis. To ensure the correct repulsion behavior of the magnetic field lines along the domain wall, we combined two untwisted monopole-antimonopole pairs by introducing a twist between them. This configuration was achieved using the following ansatz for the scalar field
\begin{align}
    \hat{\phi}=\begin{pmatrix}
    -\sin (\theta_1 - \bar{\theta}_1 + \theta_2 - \bar{\theta}_2)\sin \phi\\
    \sin (\theta_1 - \bar{\theta}_1 + \theta_2 - \bar{\theta}_2)\cos \phi\\
    -\cos (\theta_1 - \bar{\theta}_1 + \theta_2 - \bar{\theta}_2)
    \end{pmatrix},
\end{align}
where $\theta_ 1$ and $\theta_2$ ($\bar{\theta}_1$ and $\bar{\theta}_2$) correspond to the angles between the $z$-axis and the position vectors stemming from the monopoles (antimonopoles).

\begin{figure*}
    \includegraphics[trim=30 0 60 5,clip,width=\textwidth]{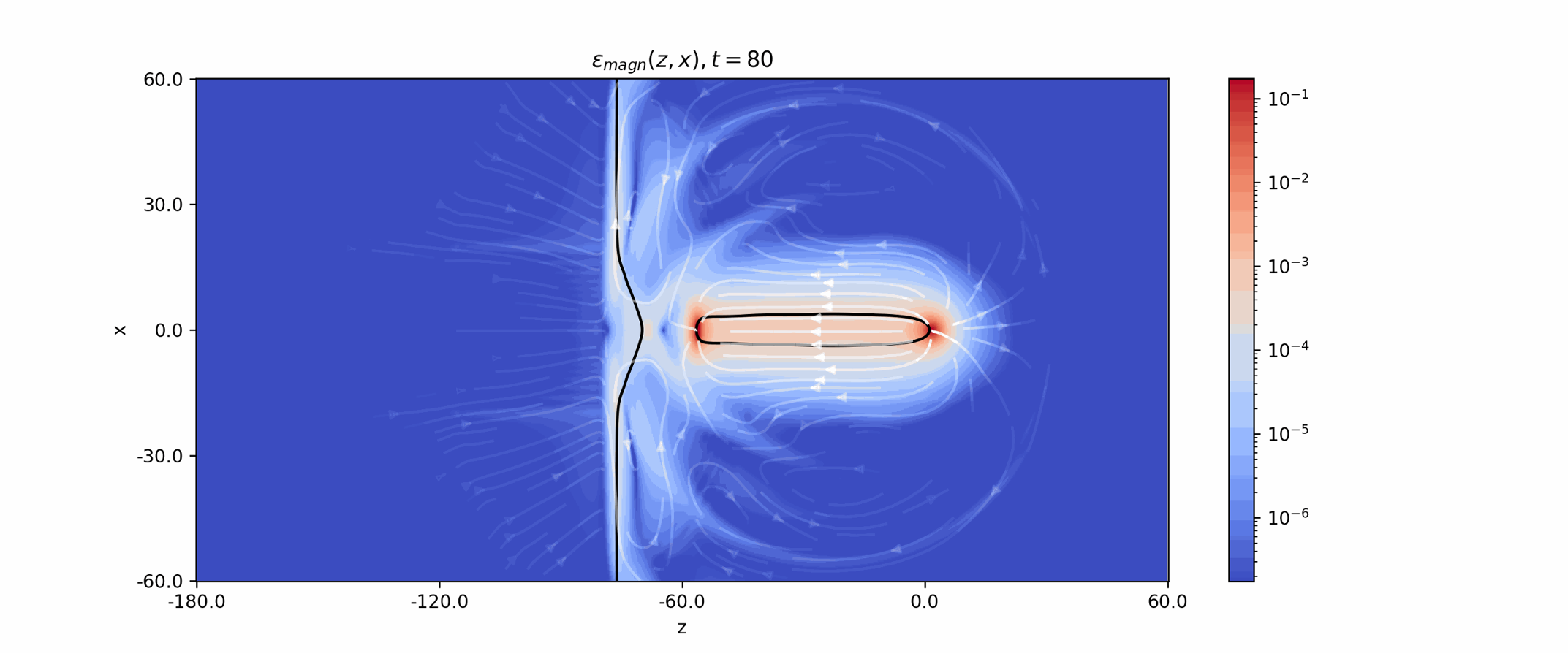}
    \begin{flushleft}
    \caption{The magnetic energy density and magnetic field for two magnetic monopoles entering the confined phase. To the units, the same applies as in Fig.~\ref{fig:magnetic-energy-density}. The first monopole enters and stretches a string. Afterward, the antimonopole enters and detaches the string from the domain wall, leading to the formation of a monopole-antimonopole pair connected by a string.}
    \label{fig:two-magnetic-monopoles}
    \end{flushleft}
\end{figure*}

The subsequent implementation followed a similar approach to the previously described model. Our observations revealed that the initial monopole extended a string, and later when the antimonopole entered the string, it caused the detachment of the string from the domain wall as can be seen in Fig.~\ref{fig:two-magnetic-monopoles}. Consequently, the monopole and antimonopole were drawn together with constant acceleration until their annihilation occurred. 
The dynamics are analogous to the one described in~\cite{Dvali:2022vwh}. The energy stored in the strings connecting them is transferred into kinetic energy of the monopole-antimonopole, which turns them ultra-relativistic \footnote{In~\cite{Dvali:2021byy}, this dynamics was applied to the production of primordial black holes. In fact, for a long enough initial string, the system will find itself within its own Schwarzschild radius well before the monopole-antimonopole annihilation, therefore leading to the production of black holes.}. 

The full-time evolution for this configuration can be found in the ancillary video, which is also available in the following link:\\
\url{https://youtu.be/IPJAPjo3nSc?si}

\noindent
Monopoles connected by a string and their dynamics in this type of model have been studied in great detail in~\cite{Dvali:2022vwh}.
 
As already discussed, the string can break by the spontaneous creation of a monopole-antimonopole pair on 
its world volume. As analysed by Vilenkin~\cite{Vilenkin:1984ib}, for a classically-stable 
string this is a tunneling process with extremely
low probability.  This analysis is equally applicable also to a long string in our case 
after its formation.  

The non-trivial question on which our analysis sheds light is, how probable is the stretching of such a string 
in a monopole wall collision?  We could imagine that once the monopole collides with the 
wall, the entire energy gets converted into radiation without ever stretching a string, as this
was observed to be the case in the collision of a confined monopole-antimonopole pair~\cite{Dvali:2022vwh}.
There, in a head-on collision, the monopoles would 
never pass each other and re-create a string. Instead,
the system decayed into waves after the first collision. 
In this respect, the two setups give very different outcomes.

The reason for this difference is the following. 
First, as explained in~\cite{Dvali:2022vwh}, 
in the case of monopole-antimonopole collision, 
after they come on top of each other, 
the system completely ``forgets" about the existence of the magnetic charges.
Basically, the monopole-antimonopole completely erases one another. 
The collision also takes away some coherence, as this is typical for the processes 
of defect erasure~\cite{Dvali:1997sa,Dvali:2022rgx}. 
Correspondingly, in its further evolution, the system has no ``profit" in re-creating a highly coherent and low entropy state of monopoles connected by a string.
Such an outcome is exponentially suppressed. 
It is much more probable to decay into a highly entropic state of waves.  
This exponential suppression is generic for the transition amplitudes
into macroscopic final states of low entropy~\cite{Dvali:2020wqi,Dvali:2022vzz}. 

The situation in the present case is very different. The reason is that the
magnetic charge is conserved. Correspondingly, the system must maintain the
monopole, no matter what. The only question is the arrangement of its magnetic flux.
Since the monopole has sufficient kinetic energy for entering 
the confining phase, the system has two choices: 1) accompany the monopole with a long string; or 
2) create at least one additional monopole-antimonopole pair for breaking it apart. 
Since the pair-creation via quantum tunneling is exponentially suppressed, 
the latter process would require a hard perturbation which would force the adjoint field to vanish.    
This is not happening since the 
monopole ``sees" the wall through the change of the expectation value 
of the fundamental field, which is a rather soft perturbation.     
Thus, the system chooses the process of stretching the long string adiabatically. 
Due to this, the outcome is a slingshot effect. 
 
The string breaking could also occur via thermal fluctuations, which will be the subject of future investigation.
    \section{Gravitational Waves}
\label{sec:gravitational-waves}
The slingshot mechanism provides a novel source of gravitational waves that can be produced in the early universe. This adds to the 
list of previously discussed sources of gravitational waves from various types of defects, such as 
colliding bubbles~\cite{Kosowsky:1991ua, Kamionkowski:1993fg}, monopole-antimonopole pairs confined by strings~\cite{Martin:1996cp, Dvali:2022vwh}, cosmic string loops~\cite{Vachaspati:1984gt}, etc. 

The interesting novelty of the slingshot source of gravity waves is that it is expected to be rather generic in a grand unified phase transition, as such transition often proceeds with the formation of domain walls separating the phases of confined and free monopoles. For example, already the minimal grand unified theory with Higgs fields in the adjoint $24_{\rm H}$ and fundamental $5_{\rm H}$ representations, allows for the coexisting temporary phases such as $SU(4)\times U(1)$ and $SU(4)$ separated by domain walls. The vacuum expectation values in these two phases have the following forms: $\langle 24_{\rm H} \rangle \propto {\rm diag}(-4,1,1,1,1),~ \langle 5_{\rm H} \rangle = 0$ and  $\langle 24_{\rm H} \rangle \propto \rm{diag}(-4,1,1,1,1),~ \langle 5_{\rm H} \rangle \propto (1,0,0,0,0)^t$ respectively. In these vacuum domains, the magnetic monopoles are in the Coulomb and confining phases respectively.
Correspondingly, the interaction between monopoles and domain walls leads to the slingshot effect. 

Of course, the purpose of the present paper is not to study the full richness of the grand unified phase portrait, which is also highly model-dependent. It suffices to notice that the slingshot can even be a dominant source of gravitational waves. In order to understand this, we can think about a single spherically symmetric expanding bubble separating the two phases. Sweeping away the monopoles by a slingshot mechanism produces gravity waves even in the absence of bubble collisions with other bubbles.
For this reason, we focus on the generic aspects of the gravitational waves produced by the slingshot, using a simple prototype example presented in previous sections. 

Notice that in our simulation, we ignore the gravitational backreaction on the dynamics of the source. 
Namely, we are assuming that the wall/string/monopole dynamics is dominated by the string tension. 
This is a legitimate assumption, provided that the tensions are below the Planck mass. 

As was shown long ago~\cite{Vilenkin1983, Ipser:1983db}, the planar infinite wall has repulsive gravity. It acts on a pointlike source of positive mass $m$ with a repulsive linear potential, given by $V(r) \sim G\,  \sigma_{\rm DW}\, m\, r$, 
where $\sigma_{\rm DW}$ is the wall tension. 
In the present case, this repulsion can compensate or even overtake the attractive potential 
due to a string. It can also prevent the monopole from crossing the wall. Notice that under the condition $\mu \gg G\, \sigma_{\rm DW}\, M_{\rm M}$ the slingshot dynamics is negligibly affected by the gravitational field of the domain wall which we assume throughout this work.  

The radiated energy at frequency $\omega$ per unit frequency and per solid angle, in direction $\vb{\hat{k}}$ ($\abs{\vb{k}}=\omega$) can be calculated by Weinberg's formula~\cite{Weinberg:1972kfs} (following the conventions of~\cite{Maggiore:2007ulw})
\begin{align}
\label{eq:weinberg-formula}
    \frac{{\rm d}E}{{\rm d}\Omega \,{\rm d}\omega}=\frac{G\, \omega^2}{2\pi^2} \Lambda_{ij,lm}(\vb{\hat{k}})T^{ij \ast}(\vb{k},\omega)T^{lm}(\vb{k},\omega)\,,
\end{align}
where $\rm{d} \Omega$ is the differential solid angle and the Fourier transform of the energy-momentum tensor is given by
\begin{align}
    T_{\mu\nu}(\vb{k},\omega)=\int_{I_t} \dd t \int_{V} \dd^3 x\ e^{i \omega t-i \vb{k}\cdot \vb{x}}\ T_{\mu\nu}(\vb{x},t)\,,
\end{align}
with $I_t$ and $V$ being the analyzed time interval and volume, respectively. These were chosen around the time and length scales of the dynamics of interest. 
The former corresponds to the duration of the source $T\simeq 80 m_{v_\phi}^{-1}$, while the latter is given by the volume spanned by the system during its evolution. Given the relativistic motion involved, $V\simeq T^3$ proved to be an optimal choice.

The operator $\Lambda_{ij,lm}$ projects a tensor into its transverse traceless part and is defined as
\begin{align}
    \Lambda_{ij,lm}(\vb{\hat{k}})\equiv P_{il}(\vb{\hat{k}})P_{jm}(\vb{\hat{k}})-\frac{1}{2}P_{ij}(\vb{\hat{k}})P_{lm}(\vb{\hat{k}})\,,
\end{align}
where $P_{ij}(\vb{\hat{k}})=\delta_{ij}-\hat{k}_i \hat{k}_j$ are projectors into the orthogonal direction of $\vb{\hat{k}}$.

In the derivation of the equation \eqref{eq:weinberg-formula}, the divergenceless condition in momentum space $k_\mu T^{\mu\nu}=0$ was assumed. The Fourier-transformed data from the numerical simulations matches this condition well; thus, we can apply formula \eqref{eq:weinberg-formula}.

\begin{figure}
  \includegraphics[width=\linewidth]{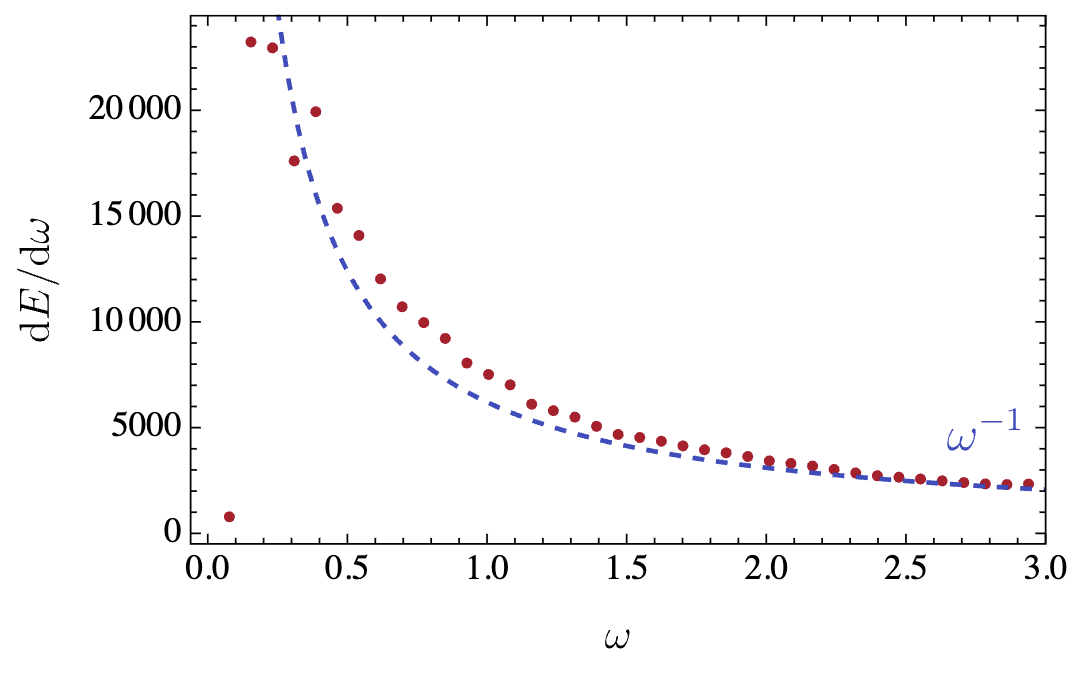}
  \caption{The energy spectrum for the slingshot effect (red points). The blue-dashed curve shows the scaling $\omega^{-1}$ for comparison.}
  \label{fig:gravitational-energy-spectrum}
\end{figure}

Since we are working on a lattice with a finite resolution and the initial configuration is an approximation (for example, the initial boost of the monopole and the domain walls leads to fictitious sources), the presence of noise in the gravitational energy spectrum, stemming from numerical fluctuations, is anticipated. Moreover, both the domain wall and the monopole are accelerated to relativistic velocities, which introduces an extra source of background in the simulation due to the finiteness of the lattice spacing. This imposes limitations on the available parameter space.

To ensure that such effects do not invalidate our analysis and that we capture only the gravitational wave signal due to the slingshot, we execute a Lorentz boost on the monopole in the opposing direction. With this strategy, we avoid a collision with the domain wall in the considered time interval and we can extract the magnitude of the background noise.

We observe that for $m_{v_\psi}=0.6 m_{v_\phi}$ \footnote{In the following $m_{v_\phi}=1$ is used, and all dimensionful quantities are expressed in units of it.} (all the other parameters are kept unchanged), the background noise in the energy spectrum is negligibly small - below $5\%$ of the energy extracted in the presence of a slingshot. In this case, the length of the string is comparable to the size of the magnetic monopole. Exploration of alternative $m_{v_\psi}$ values reveals that numerical spurious effects stops being negligible for $m_{v_\psi}\lesssim 0.4\,m_{v_\phi}$. Moreover, for $m_{v_\psi}\gtrsim 0.7 m_{v_\phi}$ the Lorenz gauge condition starts being numerically violated by more than $10\%$. 

The resulting energy spectrum, obtained upon integration over $\rm d \Omega$ is shown in Fig.~\ref{fig:gravitational-energy-spectrum}, where we fixed the Newton constant $G=1$ for simplicity\footnote{Note that the instantaneously radiated power, according to the notation of~\cite{Weinberg:1972kfs}, can be obtained by multiplying $\dv{E}{\omega}$ by $\frac{2\pi}{T^2}$.}. 

The energy spectrum is well characterized by the following scaling
\begin{equation}
\label{eq:descaling}
    \dv{E}{\omega}\propto \omega^{-1}\,.
\end{equation}
This is exemplified by the dashed blue line in Fig.~\ref{fig:gravitational-energy-spectrum}. Unfortunately, the finiteness of our numerical simulations does not permit a clear characterization at higher frequencies. However, for sufficiently high $\omega$ we expect the amplitude to be exponentially suppressed. 

\begin{figure}
  \includegraphics[trim=27 0 50 0,clip,width=\linewidth]{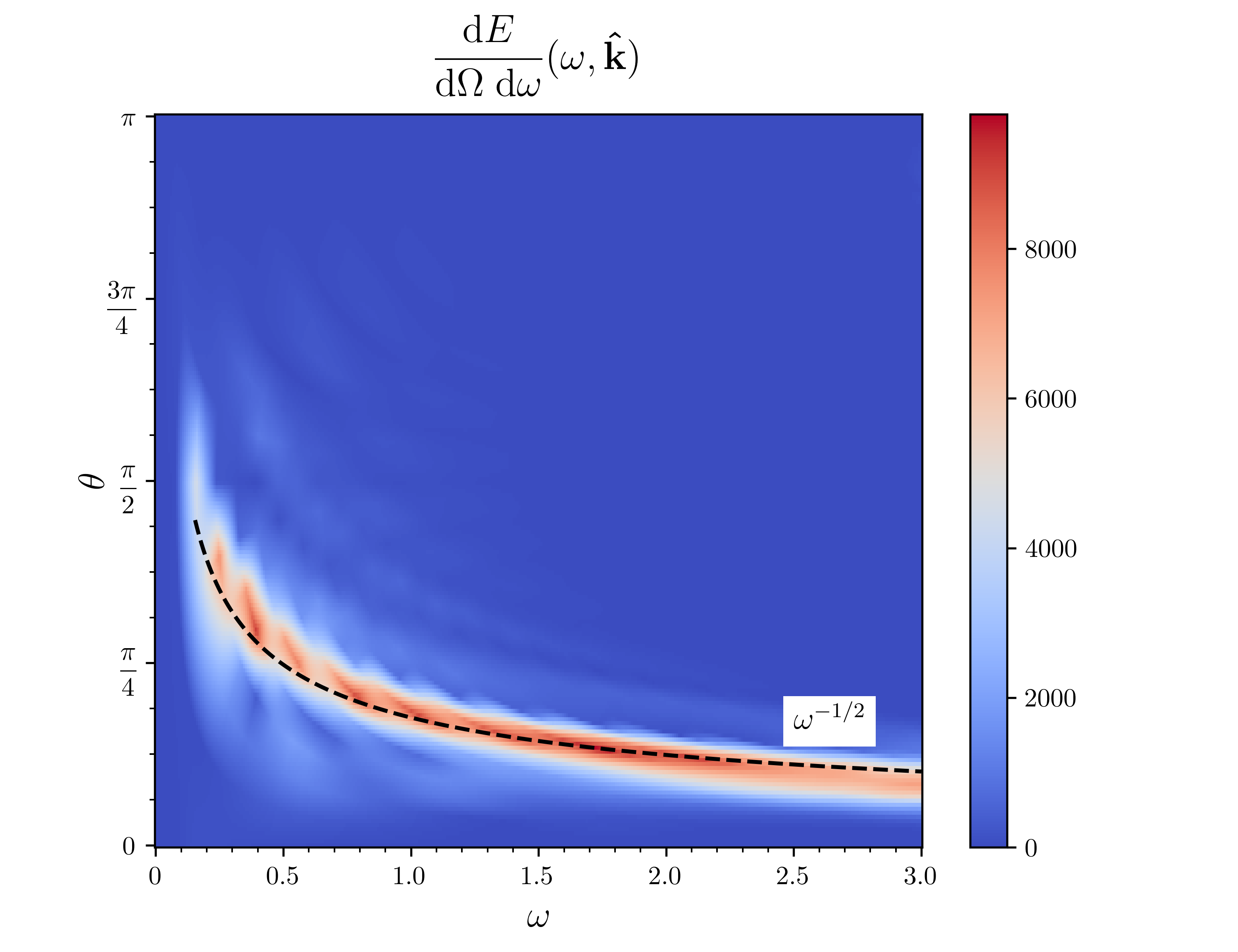}
  \caption{Angular dependence of the radiated gravitational energy as a function of $\theta$ and $\omega$. The parameters chosen are the same as outlined in the text. The angle $\theta$ gives the direction of radiation with respect to the acceleration axis. $\theta=0$ corresponds to the direction of acceleration (to the left).}
  \label{fig:radiationangular}
\end{figure}

The direction of the emission is towards the bubble wall, as seen in Fig.~\ref{fig:radiationangular}. Therein equation~\eqref{eq:weinberg-formula} is shown as a function of the axial angle $\theta$, measured from the acceleration axis, and the frequency $\omega$. As it can be seen, most of the radiation takes place in the direction of acceleration. In particular, radiation is emitted in a beaming angle with frequency dependence roughly approximated by
\begin{equation}
\label{eq:deangularscaling}
    \theta \propto \omega^{-1/2}\,,
\end{equation}
depicted by the dashed black line in the plot. 

In order to verify that the scalings \eqref{eq:descaling} and \eqref{eq:deangularscaling} are due to the monopole being accelerated by the flux tube attached to the domain wall, we performed a separate analysis in which we isolated the slingshot dynamics from the initial collision between the monopole and the domain wall. We found that indeed the main contribution to the signal in Fig.~\ref{fig:gravitational-energy-spectrum} is due to the former. 

The gravitational radiation due to the slingshot mechanism bears a close resemblance to the one emitted by a confined pair of monopole-antimonopole. In fact, also in that case the source of gravitational waves is due to monopoles being accelerated by the flux tube. As shown by Ref.~\cite{Martin:1996cp} in the limit of point-like monopoles and the zero thickness string, \eqref{eq:descaling} holds true also for that system. The result of Martin and Vilenkin was confirmed by some of us for the case of fully-resolved confined $SU(2)$ 't Hooft-Polyakov monopoles~\cite{Dvali:2022vwh}. The angular emission in the point-like limit was instead found to scale according to \eqref{eq:deangularscaling} by~\cite{Leblond:2009fq}.
 
The emitted instantaneous power for a confined monopole-antimonopole pair is given by $P\sim G\,\Lambda^4$~\cite{Martin:1996cp}. While in our numerical simulation, we have little leverage on the string tension $\mu=\Lambda^2$, we observe that the gravitational signal amplitude is roughly compatible with this scaling as we changed the value of $m_{v_\psi}$. Moreover, we observe that for low enough $m_{v_\psi}$ the radiation from the impact could become comparable to the one from the slingshot for a sufficiently short stretching of string. 

In our analysis, we focus on monopoles as the representative objects on which the confined flux can terminate. However, as we discuss below, the current analysis is general and applies also to the case of confined heavy quarks connected by gauge strings. For this latter case, the signal is produced in the regime $M \gtrsim \Lambda$, with $\Lambda\sim \sqrt{\mu_{\rm string}}$ being the confinement scale, and $M$ being the mass of the monopole or quark.
In particular, we showed that the energy spectrum decays as $\omega^{-1}$ and that the beaming angle of emission displays a $\omega^{-1/2}$ behavior in the considered parameter space.  
Therefore, a phase transition between an unconfined and confined phase can provide a specific signal in the form of gravitational waves
coming from a slingshot effect.
    \section{Implications for QCD} 

Our analysis has direct implications for QCD-like gauge theories with coexisting domains with confined and deconfined phases. Such a system was originally considered in \cite{Dvali:1996xe}. This setup possesses a domain wall on which the $SU(2)$ gauge theory is deconfined. Later in \cite{Dvali:2007nm} the domain wall was replaced by the vacuum layer the width of which can be arbitrarily adjusted.  This is the setup we shall consider now. The Lagrangian has the following form,
\begin{align} 
\label{eq:model2}
    \mathcal{L}=&-\frac{1}{2}\Tr \left(G^{\mu\nu}G_{\mu\nu}\right)+\Tr \left((D_\mu \phi)^{\dagger}\, (D^\mu \phi)\right) -U(\phi)\nonumber\\
    &+ i\bar{Q} \gamma^{\mu} D_\mu Q -  M_Q \bar{Q}Q\,,
\end{align}
where $\phi$ is a Higgs field in the adjoint representation of the 
$SU(2)$ gauge symmetry with the potential 
\begin{align}
    U(\phi)=\lambda \Tr \left(\phi^2\right)\left(\Tr \left(\phi^2\right)-\frac{v_\phi^2}{2}\right)^2.
\end{align}
The $SU(2)$ gauge sector of the theory, as well
as the corresponding notations, are the same as in previous examples.
The fermion content of the theory consists of (for simplicity) a single flavor of a heavy quark $Q$, in the fundamental representation of $SU(2)$. 
Under ``heavy" we mean that the mass of the quark $M_Q$ is above 
the confinement scale of the $SU(2)$ theory, which we denote by $\Lambda$. 

The potential $U(\phi)$ possesses the following two vacua. 
In the first vacuum $\phi = 0$, the perturbative spectrum of the theory consists of an adjoint scalar of mass $m_{\phi} = \lambda v_{\phi}$, the fundamental quark of mass $M_Q$, and a massless $SU(2)$ gauge field. The effective low energy theory is therefore a massless Yang-Mills.  
As it is well-known, this theory becomes confining and generates a mass gap at the corresponding QCD scale $\Lambda$.
Correspondingly, an electric flux of gluons confines into flux tubes 
that represent QCD strings~\cite{tHooft:1973alw, tHooft:1974pqx} with tension $\mu_{\rm string}\sim \Lambda^2$. The lowest mass excitations about this vacuum are colorless glueballs, which can be thought of as closed QCD strings. 
The spectrum also includes mesons, which represent quark-antiquark pairs 
connected by flux tubes and open strings.
  
The effect of the adjoint scalar $\phi$ on the confinement can be consistently ignored for $\Lambda \ll m_{\phi}$, which we assume for definiteness.  

In the second vacuum, classically, we have $\phi = v_\phi$, and thus the $SU(2)$ gauge group is Higgsed down to the $U(1)$ subgroup. The bosonic spectrum of the theory consists of a real $U(1)$ neutral scalar of mass $m_{\phi} = \sqrt{\lambda} v_\phi^2$, a charged (complex) massive gauge boson of mass $m_{v_\phi} = g v_\phi$, and a massless Abelian $U(1)$ gauge field.
In addition, of course, there exists a massive fermion $Q$. 
For $m_{v_\phi},m_\phi \gg \Lambda$, the quantum effects from the massive modes can be safely ignored and the effective low-energy theory consists of a $U(1)$ gauge theory in the Coulomb phase. 

The theory possesses a domain wall solution separating the two phases. Classically the solution can be found exactly. In the above approximation, the quantum effects of the shape of the solution are small, but they can lift the degeneracy of the two vacua. The bias can create a pressure difference which accelerates the wall. This does not change much in our discussion, and in fact, the controlled acceleration of the wall can be welcome for the study of the scattering as we have seen in the numerical simulations. The bias can be controlled by proper adjustment of the parameters. 
   
The long-distance physical effects of the heavy quark in the two domains are different. In the $U(1)$ domain, the quark produces a $U(1)$ Coulomb electric field. 
In the $SU(2)$ domain, the quark is a source of the flux tube. This flux tube 
can either terminate on an antiquark or on the wall. In the latter case, 
the QCD-electric flux flowing through the tube opens up in the form of the $U(1)$ Coulomb flux on the other side of the wall. 

Notice that a long QCD string can break by nucleation of quark-antiquark pairs.  However, this process is exponentially suppressed for $M_Q > \Lambda$, 
and the string can be stable for all practical purposes.  This suppression is similar to the exponential suppression of the decay of a magnetic Nielsen-Olesen string via nucleation of monopole-antimonopole pairs. 
In what follows we shall assume the regime of such stability.  
   
This structure makes it clear that in certain aspects the model \eqref{eq:model2} represents an electric ``dual" of the previously discussed model \eqref{eq:model1}. 
The role of monopoles is played by the heavy quarks, whereas the role of the magnetic field is taken up by the electric flux of QCD. In both cases, the wall separates the confining and Coulomb phases for the given flux. 

The problem of monopole scattering at the wall is mapped on the scattering between the wall and a heavy quark. 
When a quark moves across the wall from the $U(1)$ phase to the confining one the flux carried by it stretches in the form 
of the string that opens up as a Coulomb flux at the entry point. By conservation of the $U(1)$ flux, the charge measured by an observer in the Coulomb vacuum must exactly match the $U(1)$ charge of the initial quark.  

However, this conservation can be fulfilled in two ways. Upon entry into the confining vacuum, the string may stretch without breaking up leaving the initial quark as the source of the flux. 
Alternatively, the string may break up by nucleating additional quark-antiquark pairs or closed strings. That is the system can transfer most of its initial energy into mesons and glueballs. 

The process is very similar to the scattering of a magnetic monopole at the wall. In that case, we saw that the string never breaks up. 
If the analogy can be trusted, we would conclude that the same must be true for the case of a quark entering the confinement domain from the Coulomb one.  

For $M_Q \gg \Lambda$ such a behavior is relatively easy to justify. The two factors are defined: 1) The continuous memory of the initial state; and 2) the softness of the process.  

First, notice that by the gauge invariance, the $U(1)$ charge is 
fully conserved. The generation of the mass gap in the confining domain 
does not affect this conservation. Due to this, a charge was placed in the 
$U(1)$ Coulomb domain and never creates any image charges on the confining side
and the flux is repelled without any screening~\cite{Dvali:1996xe,Dvali:2007nm}.
This is the key to the localization of a massless photon by the mechanism of~\cite{Dvali:1996xe}. 

Therefore, when the quark crosses the wall and enters the confining region, the memory of the initial state is maintained in the form of the Coulomb electric flux. 
The total flux is conserved and is exactly equal to the 
$U(1)$ charge carried by the quark. This flux can be monitored 
by measuring it on the Coulomb side of the wall.   
Now, when the free heavy quark enters the confining region, no hard collision takes place. The Coulomb-electric flux carried by the quark gathers into a tube of the thickness 
$\Lambda^{-1}$, which is much larger than the de Broglie and Compton wavelengths of the quark. Correspondingly, the dynamics are 
soft and no processes with momentum transfer 
exceeding the quark mass take place. Correspondingly, the probability of quark-antiquark pair creation is exponentially suppressed. This results in a slingshot effect during which  
a long thick string is stretched in the wake of the quark.
The resulting deceleration process is soft. 
The decay of a formed long string via pair creation is 
exponentially suppressed due to the usual reasons. 

This reasoning must remain applicable also for the lower masses of quarks that are closer to the QCD scale, $M_Q \gtrsim \Lambda$, as long as the exponential suppression of the string decay is maintained. 
That is, provided the parameters are such that the static long string is stable against the breakup via pair-nucleation, the relativistic quark entering the confining domain 
is expected to exhibit a slingshot effect.   
 
Just like in the monopole case, this outcome is different from what is expected from the collision of a quark-antiquark pair connected by a string. In this case, due to the absence of a net $U(1)$ charge, upon annihilation of quarks, 
the memory about the pre-existing charge dipole is gone. The system then chooses to hadronize in a multiplicity of glueballs and mesons rather than to stretch a long string.

Apart from its quantum field theoretic importance,  
the slingshot effect with quarks can have equally interesting cosmological implications, since the coexistence of confined 
and deconfined phases are generic in the cosmological evolution of various extensions of the standard model, such as grand unification. The quark slingshot effect can supplement the mechanism of the primordial black hole formation proposed in~\cite{Dvali:2021byy}. Now, instead of quarks connected by a string, the black hole can form by smashing a highly energetic quark accelerated by a slingshot into a wall. In addition, the quark slingshot effect can be the source of gravitational waves in a way very similar to the monopole slingshot case discussed in the previous chapter. 
    \section{Slingshot of Confined Vortexes and strings}

The slingshot effect is not limited to confined point-like sources, such as monopoles or quarks with attached strings. Both are objects of co-dimension $3$ confined by a connector of co-dimension $2$ (string). Objects of different
co-dimensionality can exhibit the slingshot effect.  In general, sources of co-dimension $d$ are confined by co-dimension $d-1$ agents. 
For example, in $3+1$ dimensions, strings that have co-dimension $2$ can be confined by domain walls which are co-dimension $1$ objects.
A well-known example of such confinement is provided by strings bounding domain walls that stretch between them~\cite{Kibble:1982dd, Vilenkin:1982ks}.   
Similarly, in $2+1$ dimensions, vortices can be confined by strings~\cite{Dvali:1991ka}.

In the current section, we shall study the slingshot effect for this case. For this, we shall extend the model of confined vortices
in $2+1$ dimensions (strings in $3+1$ dimensions) introduced in~\cite{Dvali:1991ka}, by allowing an additional vacuum in which vortices (strings) are not confined. The $2+1$-dimensional model of this sort has double usefulness as on the one hand it captures the dynamics of the string slingshot in $3+1$ dimensions and on the other hand it represents  
a toy version of the monopole slingshot discussed in the previous sections. 

The key concept of the model involves replacing the adjoint and fundamental scalar fields with complex scalar fields of different charges under an Abelian symmetry. Instead of an $SU(2)$ gauge model, we consider a $U(1)$ gauge theory. The symmetry-breaking mechanism involves 
two scalar fields. The first scalar field of charge $q_\phi=g$, denoted as $\phi$, is breaking the $U(1)$ symmetry down to $Z_2$ symmetry. 
This discrete symmetry is broken further by a second scalar field of charge $q_{\chi} =\frac{g}{2}$, referred to as $\rchi$.
This field changes the sign under the $Z_2$ transformation.  

The Lagrangian governing this model is expressed as follows~\cite{Dvali:1991ka}
\begin{align}
    \mathcal{L}=&(D_\mu \phi)^\ast (D^\mu \phi)+(D_\mu \rchi)^\ast (D^\mu \rchi)\nonumber\\
    &-\frac{1}{4}F_{\mu\nu}F^{\mu\nu}-U(\phi,\rchi)\,,
\end{align}
with the potential
\begin{align}
    U(\phi,\chi)=&\lambda_\phi (\abs{\phi}^2 -v_\phi^2)^2+\lambda_\chi (\abs{\rchi}^2 -v_\chi^2)^2 \abs{\rchi}^2\nonumber\\
    &+\beta \phi^\ast \rchi^2 +c.c.\,.
\end{align}
The covariant derivatives and the field strength tensor are given by
\begin{align}
    F_{\mu\nu}&=\partial_\mu A_\nu -\partial_\nu A_\mu\,,\\
    D_\mu \phi &= \partial_\mu \phi +i g A_\mu \phi\,,\\
    D_\mu \rchi &= \partial_\mu \rchi +i \frac{g}{2} A_\mu \rchi\,.
\end{align}

Again, the novelty as compared to~\cite{Dvali:1991ka} is that 
the potential for the $\rchi$ field is designed in such a way that in addition to the $Z_2$-Higgsed phase, in which both fields have non-zero vacuum expectation values, there coexists a $Z_2$-invariant phase in which 
the $\rchi$ field vanishes. This is possible as long as 
$\abs{\beta}$ is sufficiently small. Namely, if
$|2\beta v_\phi| < |\lambda_\chi v_\chi^4|$. 
Notice, that for simplicity we have omitted the phase-independent interaction terms such as $|\phi|^2|\rchi|^2$. 
Such terms do not play any role in the confinement of vortices. 
The crucial term in this respect is the phase-dependent 
interaction term with the coefficient $\beta$. This term defines the relative charges of the two fields. 

Let us now discuss the properties of vortices in these two vacua. 
In the $Z_2$-invariant vacuum, only the $\phi$ field 
has a non-zero vacuum expectation value. Its absolute value is constrained to the field and is $\langle |\phi| \rangle = v_{\phi}$, whereas the phase degree of freedom $\theta_{\phi}$
becomes the longitudinal component of a massive vector field through the usual Higgs effect. 

Correspondingly, the spectrum of the theory contains a Nielsen-Olesen vortex solution, given by the ansatz~\cite{Nielsen:1973cs}
\begin{align}
    A_i (r,\theta)&=\frac{n}{g}\varepsilon_{ij} \frac{r^j}{r^2} K(r)\,,\\
    \phi (r, \theta)&=v_\phi e^{in\theta} H(r)\,,
\end{align}
where $n$ is the winding number, and $K(r)$ and $H(r)$ are the profile functions that we found again by a numerical relaxation method by solving the following differential equations
\begin{align}
    K''&=\frac{K'}{r}-m_{v_\phi}^2 H^2 (1-K)\,,\\
    H''&=-\frac{H'}{r}+\frac{(1-K)^2}{r^2}n^2 H+\frac{m_{h_\phi}^2}{2}H (H^2-1)\,.
\end{align}

In the $Z_2$-Higgsed phase also the $\rchi$ field gets 
a non-zero vacuum expectation value. 
For a small enough $\beta$-term, its absolute 
value is approximately equal to $\langle |\rchi| \rangle \simeq v_{\chi}$. 
Due to this, the gauge field receives a further mass contribution $m_{v_\chi}= v_\chi g/\sqrt{2}$. The Higgs masses are approximately given by $m_{h_\phi}=2\sqrt{\lambda_\phi} v_\phi$ and $m_{h_\chi}=2\sqrt{\lambda_\chi}v_\chi^2$.

The further breaking of the $Z_2$ symmetry by the vacuum expectation value of $\rchi$ puts the $\phi$ vortices in the confining phase~\cite{Dvali:1991ka}. 
The dominant effect is due to the interaction $\beta$-term, which is phase-dependent. Notice that without this term, 
no confinement would occur. 

The reason is the following. For $\beta =0$, 
the theory would be invariant under two independent global symmetries $U(1)_{\chi}\times U(1)_{\phi}$ with only one subgroup being gauged. 
The gauged subgroup leaves the following combination of the phases invariant, 
\begin{equation} \label{INV}
    \Theta \equiv \theta_{\phi} -2\theta_{\chi} \,.
\end{equation}
This gauge-invariant phase shifts under the additional global $U(1)$ symmetry which emerges for $\beta =0$. 
The breaking of this symmetry by the combination of the two vacuum expectation values results in the emergence of a massless Goldstone boson. In the regime $v_{\phi} \gg v_{\chi}$, this would-be-Goldstone boson resides mostly in the phase $\theta_{\chi}$. 
  
Correspondingly, for $\beta =0$, the vacuum expectation value of $\rchi$ would lead to the formation of a second type vortex. 
Around each vortex, the two phases can in general have independent winding numbers. 

Notice that some vortices would be ``semi-global" ~\cite{Dvali:1994qf}. 
In particular, a vortex around which both fields have unit winding numbers would have a logarithmically divergent gradient energy since the gauge field would be unable to compensate the winding of both phases simultaneously due to the difference in their gauge charges. 

We are interested in the regime of confined vortices which takes place for $\beta \neq 0$.
 First notice that, since the $\beta$-term explicitly breaks the global $U(1)$ symmetry, the would-be-Goldstone degree of freedom gets the mass 
 \begin{equation} \label{GoldM}
     m_g^2 \simeq \abs{4\beta v_{\phi}} \,.
 \end{equation}
Minimization of the $\beta$-term term forces the alignment in the phases of the $\phi$ and $\rchi$ fields. For $\beta < 0$, 
the term is minimized for
\begin{equation} \label{eq:locked}
    \theta_{\phi} = 2\theta_{\chi} \,.      
\end{equation}

However, such a relationship cannot be maintained everywhere 
around the $\phi$ vortex with winding number one around which the phase shift is $\Delta \theta_{\phi} = 2\pi$. In the light 
of \eqref{eq:locked}, this would imply that the corresponding change of the phase of $\rchi$ around a closed path is $\Delta \theta_{\phi} = \pi$, which violates a single-valuedness of the vacuum expectation values. 

To avoid the conflict, the field compromises:
The presence of the $\beta$-term makes sure that around the closed contour enclosing the $\phi$ vortex the phase of $\rchi$ experiences a jump (rapid change) from $\pi$ to $2\pi$ within a region of thickness $\sim m_g^{-1}$.  This region represents a string 
that is attached to the $\phi$ vortex. 

Far away from the vortex core, the corresponding configuration
for the gauge invariant combination of the two phases \eqref{INV} 
can be found by solving the sine Gordon equation, 
\begin{equation}
    \Theta'' -  m_g^2 \sin (\Theta) = 0\,,  
\end{equation}
where the derivative is taken with respect to a perpendicular coordinate $y$. This equation has a well-known solution, 
\begin{equation}
\label{eq:solution-sine-Gordon}
    \Theta(y) = 4 \tan^{-1}(e^{m_gy})\,,  
\end{equation}
which interpolates from $\Theta =0$ to $\Theta = 2\pi$. 

In the vacuum with broken $Z_2$ symmetry, the string can terminate on another vortex or an antivortex, and the two get confined.
In the present case, we have a separate domain with unbroken $Z_2$ symmetry. This gives a possibility for the $Z_2$ string to terminate on a domain wall separating the two phases.  
 
The domains with free and confined $\phi$ vortices are separated by 
a domain wall in which the $\rchi$ field interpolates from 
$0$ to $v_{\chi}$.
This domain wall solution can be found by fixing the $U(1)$ direction, e.g. $\xi = \Re \rchi$ and solving the Bogomolny equation~\cite{Bogomolny:1975de}
\begin{align}
    \xi(x)=\frac{v_\chi}{\sqrt{1+e^{m_{h_\chi}x}}}\,.
\end{align}
This domain wall separates the $Z_2$-invariant phase from the $Z_2$-Higgsed phase.\\

We now turn to the analysis of a slingshot effect experienced 
by a $\phi$ vortex that passes from the $Z_2$-invariant domain into the Higgsed $Z_2$ domain. Just like in the case of a monopole, the vortex stretches a string that connects it to the boundary of the two phases. 

Since the gauge field is massive in both regions, it has no long-range effects. Its influence vanishes exponentially at distances larger than the Compton wavelength of the photon. 
Consequently, a detailed study of its behavior in close proximity to the domain wall is unnecessary, provided we assume an initial configuration in which the vortex is far enough away from the wall. This implies that the ansatz for the $\phi$ and $A_\mu$ fields in the numerical simulation does not require adaptation to account for the presence of the wall. 
The ansatz for the $\rchi$ field can be written as
\begin{align}
\label{eq:ansatzrchifield}
    \rchi(x,y) =\xi(x)\, e^{in \theta /2}\,,
\end{align}
where $\theta = \arctan\left(y/(x-x_0)\right)$, $x_0$ being the position of the vortex. The above choice for $\rchi$ minimizes the $\beta$ coupling in the broken $Z_2$ phase. Moreover, ansatz \eqref{eq:ansatzrchifield} ensures the single-valuedness of $\rchi$ since $\xi(x)$ is vanishing in the unbroken region.

In order to conduct the simulation, we employed the same numerical methods as described earlier. However, since our current analysis is limited to two dimensions, the axial symmetry method is not necessary. The lattice size, lattice spacing, time step, and the investigated time interval remained the same as in the magnetic monopole setup. Furthermore, the boundary conditions stay similar. Absorbing boundaries were utilized for $\phi$ and $A_\mu$, while the Dirichlet boundary condition was applied to $\rchi$ in the $x$-direction, accompanied by a periodic boundary condition in the $y$-direction. Note that the imaginary part of $\rchi$ is anti-symmetric in $y$-direction.

We set $m_{v_\phi}$ and $m_{h_\phi}$ to one and $g=1/\sqrt{2}$. Additionally, we took the following parameter values: $m_{v_\chi}=0.3 m_{v_\phi}$, $m_{h_\chi}=0.8 m_{v_\phi}$, and $\beta=-0.01 m_{v_\phi}^{3/2}$.

The initial distance between the vortex and the wall was chosen to be $d=40 m_{v_\phi}^{-1}$ and the velocities were $0.8$ and $-0.8$ for the vortex and domain wall respectively.

\begin{figure*}
    \centering
    \begin{subfigure}{0.8\textwidth}
    \includegraphics[trim=30 0 30 5,clip,width=\textwidth]{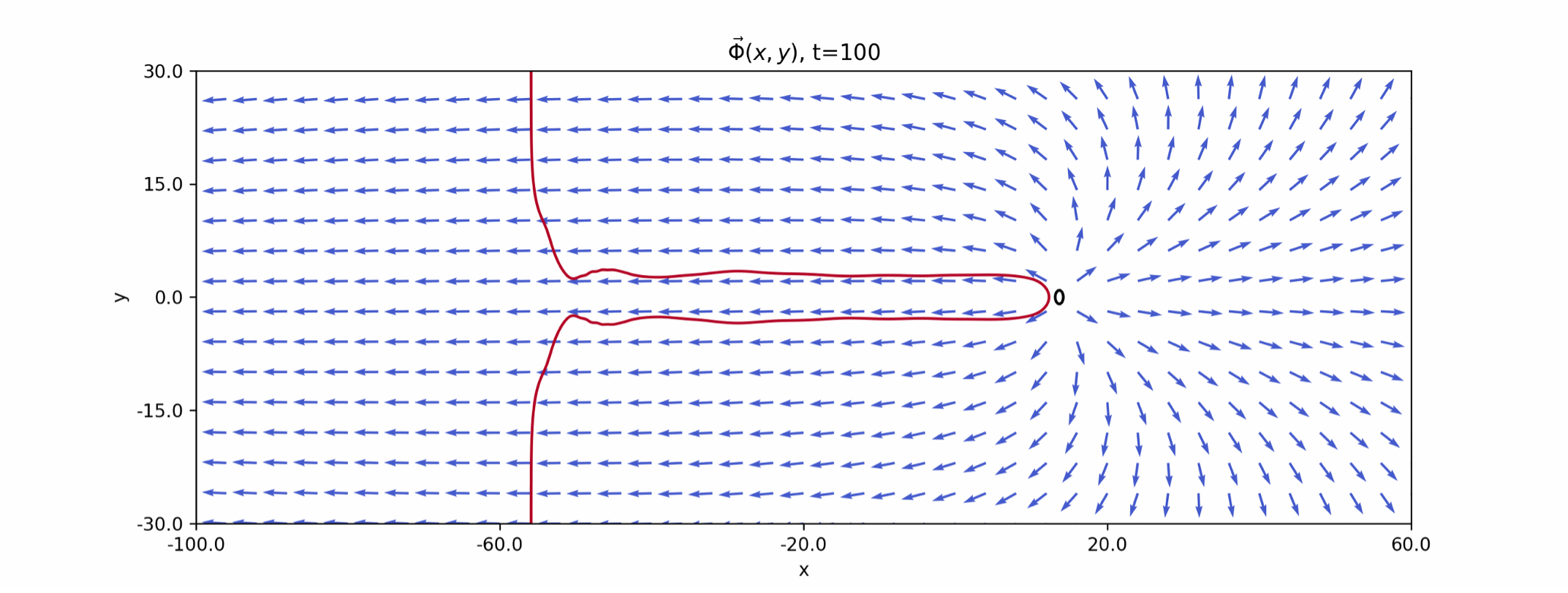}
  \end{subfigure}
  \begin{subfigure}{0.8\textwidth}
    \includegraphics[trim=30 0 30 5,clip,width=\textwidth]{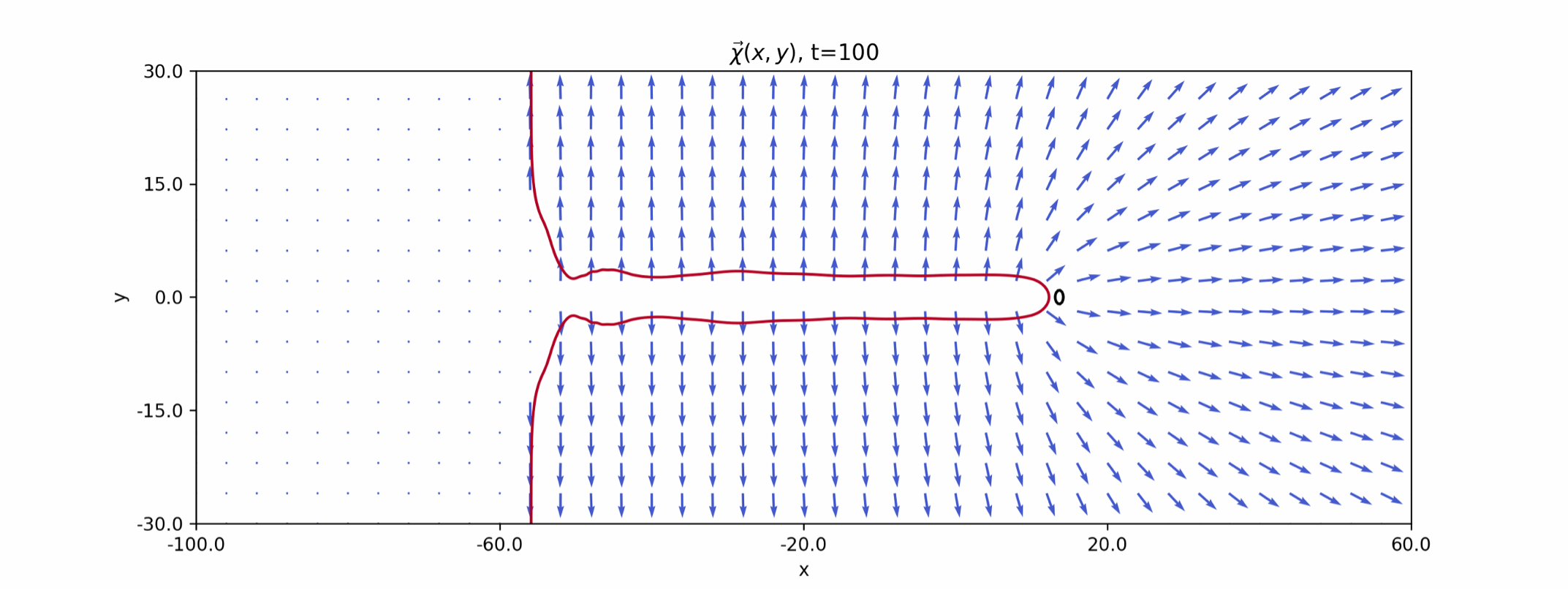}
  \end{subfigure}
    \caption{The scalar field vector $(\Re \phi, \Im \phi)^T$ (top) and $(\Re \chi, \Im \chi)^T$ (bottom) at time $t=100 m_{v_\phi}^{-1}$. The length values are given in units of $m_{v_\phi}^{-1}$. The red line represents the contour $\abs{\chi}=v_\chi/2$ and the black circle the contour $\abs{\phi}=v_\phi/2$.
    We can observe that the vortex that is entering the $Z_2$-Higgsed phase is connected to the domain wall by a $Z_2$ string.}
    \label{fig:scalar-field-2d-model}
\end{figure*}

From the simulation, we can observe that the vortex stretches a $Z_2$ string when it enters the $Z_2$-Higgsed phase as can be seen in Fig.~\ref{fig:scalar-field-2d-model}. 
The formation happens very similarly to the magnetic monopole case. The qualitative difference is that there is no magnetic flux inside the $Z_2$ string due to the short-range behavior of the vortex gauge field.

The minimization of the interaction term results in $\rchi^2$ being proportional to $\phi$. Given the vortex's winding number of one, this proportionality implies a winding of $1/2$ in the $\rchi$ field at the end of the string. Consequently, the field vectors exhibit a rotation by $\pi$ around the string's end. Within the string, the phase is changing according to equation \eqref{eq:solution-sine-Gordon}. This rotational behavior explains why the $Z_2$ string does not detach, as a rotation by $2\pi$ is necessary for the detaching to occur. Therefore, the formation of the $Z_2$ string is purely explained by topology.

Unlike in the case of monopoles 
or quarks in $3+1$ dimensions, the slingshot effect of vortices (strings) happens without the confinement of the gauge flux.
Instead, what confines within the string connecting two vortices 
is the flux of gradient of the Goldstone field which 
in the $\beta=0$ limit becomes uniformly distributed 
around the vortex resulting in $2+1$-dimensional Coulomb interaction between them. For vortices separated by a distance 
$r$ the interaction potential is  $\propto \ln(r)$. 
If $\beta \neq 0$, for distances $r \gg m_g^{-1}$, the  potential is converted into a linear confining 
potential $\propto r$.

Again, we can add an antivortex that enters the string later, leading to the breaking of the string and subsequent annihilation of the vortex-pair. However, the $2+1$-dimensional model possesses a distinctive feature not present in the magnetic monopole model. In this case, it is possible for a second vortex to enter the system instead of an antivortex. As a result, the string breaks, causing the two vortices to be drawn together until they form a bound state. This bound state exhibits a winding number of two in the $\phi$ field and a winding number of one in the $\rchi$ field.

\begin{figure*}
    \centering
    \begin{subfigure}{0.4\textwidth}
    \includegraphics[trim=0 0 0 0,clip,width=\textwidth]{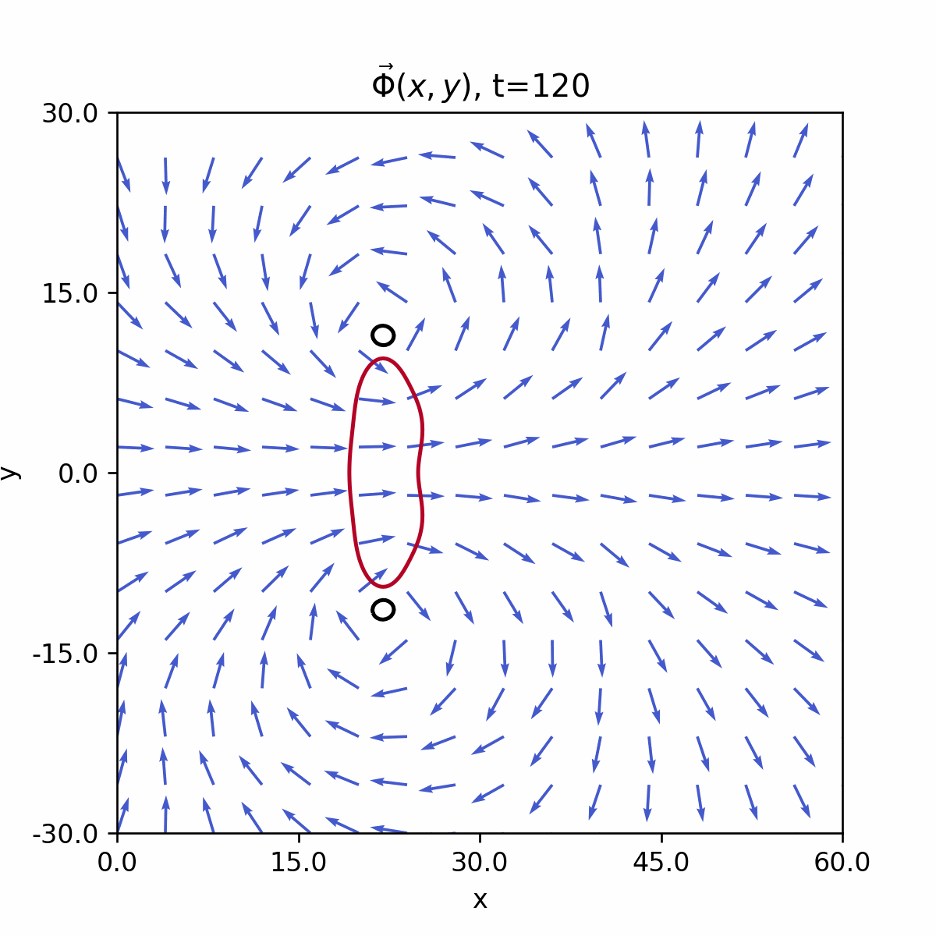}
  \end{subfigure}
  \begin{subfigure}{0.4\textwidth}
    \includegraphics[trim=0 0 0 0,clip,width=\textwidth]{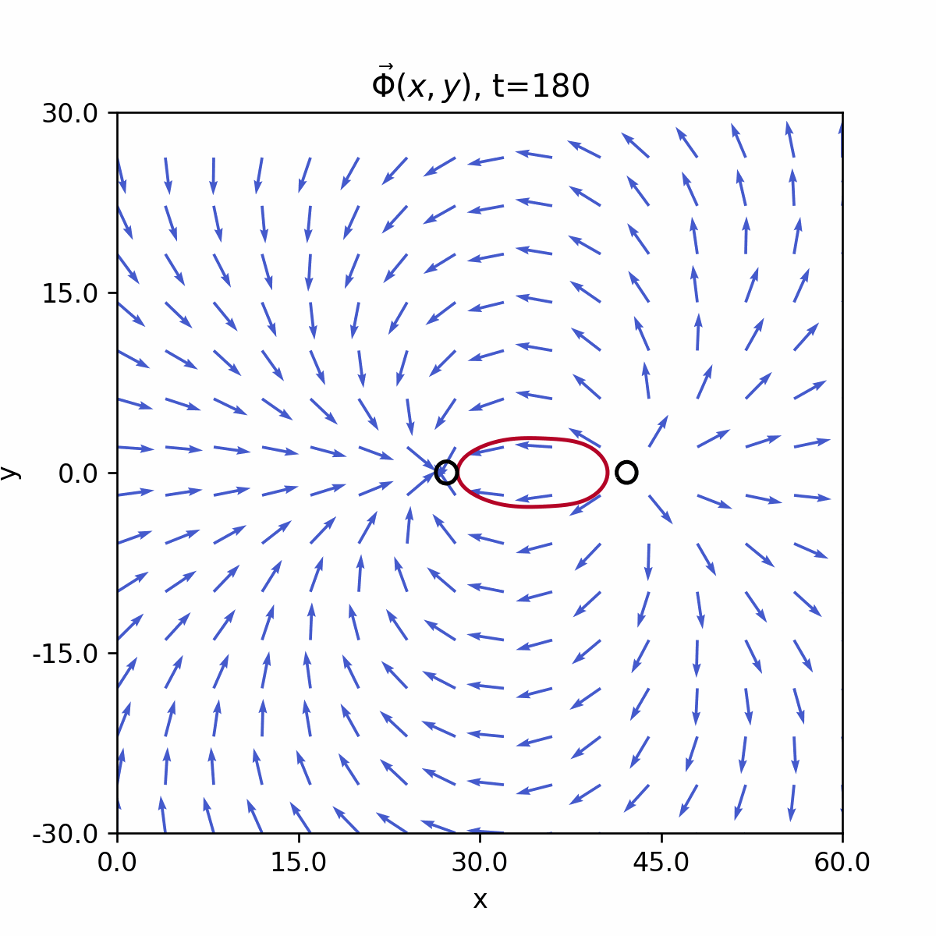}
  \end{subfigure}
    \caption{The scalar field vector $(\Re \phi, \Im \phi)^T$ at times $t=120 m_{v_\phi}^{-1}$ and $t=180 m_{v_\phi}^{-1}$. The frames show moments after two vortices of the same winding entered one after one the $Z_2$-Higgsed phase. We observe that the two vortices are connected by a string and form an oscillating bound state.}
    \label{fig:2d-model-bound-state}
\end{figure*}

During the collision, we observe that the two vortices scatter at an angle of $\pi/2$. This scattering behavior has been previously explained and analyzed using the moduli space approximation in \cite{Myers:1991yh, Shellard:1988zx}. Due to the binding effect of the $\rchi$ field on the two vortices, this right-angle scattering occurs repeatedly. In Fig.~\ref{fig:2d-model-bound-state} two moments of this bound state are illustrated.

The results of the simulations can be also found in the video attached as an ancillary file or at the following link:\\
\url{https://youtu.be/IPJAPjo3nSc?si}

The behavior observed in the $2+1$-dimensional model can be seamlessly extended to the three-dimensional case. The 
$\phi$-vortices are lifted in strings that extend in an additional dimension. Furthermore, the $Z_2$ string that in $2+1$ confines vortices, in $3+1$ is lifted into a domain wall that confines strings.  

In a manner analogous to the connected vortex-vortex pair and vortex-antivortex pair, we can now have a string-string pair and string-antistring pair connected by a domain wall.
Within the string-string scenario, the entities align to create a bound state, adopting a cable-like configuration characterized by identical right-angle oscillations as witnessed in the case of vortices. In the string-antistring case, however, they will annihilate.

Just like in the monopole/quarks case, the string/vortex slingshot effect can have cosmological implications as it is expected to take place in various extensions of the standard model. 
    \section{Conclusion and Outlook}
\label{sec:conclusion-and-outlook}
In this paper, we introduced and numerically studied the slingshot effect and its implications such as gravitational waves.

In the first example, we studied the scattering process in which a magnetic monopole crosses a domain wall separating the vacua 
of magnetic-Coulomb and magnetic-confining phases. 
The setup is achieved by variants of an $SU(2)$-symmetric model with coexisting phases of the type discussed earlier~\cite{Dvali:2002fi}. 
It possesses two vacuum states. In one of them, the $SU(2)$ is Higgsed  down to $U(1)$ and the spectrum contains 't Hooft-Polyakov monopoles. In the neighboring vacuum, the 
$U(1)$ symmetry is further Higgsed and the photon has a non-zero mass. In this vacuum, the monopoles can only exist in a confined form. The magnetic flux of the monopole is trapped in a tube, the Nielsen-Olesen string~\cite{Nielsen:1973cs}. 
The two vacua are separated by a domain wall.
   
We study a process in which the monopole with a high kinetic energy crosses over from the $U(1)$ Coulomb phase into the Higgs phase. We observe that upon entering the $U(1)$ Higgs domain, the monopole becomes connected to the wall by a long string. 
The string carries the magnetic flux of the monopole which opens up on the other side of the wall in the form of the Coulomb-magnetic field. 
  
Despite the fact that the conservation laws permit the disposal of the kinetic energy of the monopole in the form of waves, without stretching a long string,  this does not happen. 
Instead, the system creates a string that follows the monopole. 
The string tension tends to pull the monopole back towards the wall, exhibiting a sort of slingshot effect.  
    
This outcome is different from the previously  studied case~\cite{Dvali:2022vwh} of scattering of a monopole-antimonopole pair connected by a string.
In the point-like approximation, which does not resolve the structure of monopoles and strings, one cannot exclude that monopoles pass through each other, oscillating and re-stretching the string multiple times~\cite{Martin:1996cp}.
However, the simulation of the fully resolved system~\cite{Dvali:2022vwh} showed that in a head-on collision, the monopoles never pass through each other. Instead, they decay into waves. 
    
In~\cite{Dvali:2022vwh} this behavior was explained by the following factors. First when the monopole and antimonopole overlap, they effectively erase the magnetic charges of each other and the system forgets about the magnetic dipole.   
Also, as in the generic cases of the erasure of defects~\cite{Dvali:1997sa}, the coherence is lost. From this point on, the system evolves into the highest entropy configuration which is given by the waves, as opposed to monopoles connected by a long string. 
The latter configuration carries much lower entropy. 
The outcome can be interpreted as a particular case of a generic phenomenon the essence of which is an exponential suppression of the creation of the low-entropy macroscopic objects in collision processes~\cite{Dvali:2020wqi, Dvali:2022vzz}.
       
We explained that in the present case, the situation 
is very different due to the conservation of the net magnetic charge and the softness of the monopole-wall collision. At no point, the monopole encounters a phase in which the expectation value of the adjoint Higgs vanishes. Therefore, neither the coherence nor the memory of the preexisting state is lost. 
The monopole, due to its high kinetic energy, enters the confining phase softly 
and its magnetic flux stretches in the form of a string. 

We argued that similarly to the earlier discussed analogy between confined quarks and monopoles~\cite{Dvali:2022vwh}, the current behavior must also be shared by a dual QCD-like theory. 
   
In order to make the mapping more precise, as an electric dual version of the present model, we have used the construction analogous to~\cite{Dvali:2002fi, Dvali:2007nm}. This gauge theory represents the $SU(2)$ QCD which possesses two vacua. In one vacuum, the theory confines at a scale $\Lambda$, and quarks are connected by the QCD flux tubes. 
In the other vacuum, $SU(2)$ is Higgsed down to $U(1)$ and the theory is in the $U(1)$ Coulomb phase mediated by a massless photon.   
The theory possesses a domain wall separating the two phases. 
Due to the mass gap $\Lambda$ in the confining vacuum, the massless photon is repelled from there and is  localized within the Coulomb vacuum via the dual-Meissner mechanism of~\cite{Dvali:1996xe}.    
  
In analogy with the monopole case, we consider a scattering process in which an energetic heavy quark crosses over from the $U(1)$ Coulomb domain into the $SU(2)$-confining one. We argued that the same behavior is expected as in the case of a monopole in the dual theory. 
That is, upon entering the confining phase, the quark will softly stretch a long QCD string. The string transports the electric flux of the quark to the wall and spreads it out in the other domain in the form of the $U(1)$ Coulomb field.  
This should be the likely outcome as opposed to hadronizing into a high multiplicity of mesons and glueballs.  
    
Our reasoning is the same as in the monopole case. The conservation of the $U(1)$ charge forces the system 
to maintain the quark. The creation of additional quark-antiquark pairs that would break the string, requires collisions with high momentum transfer. These are absent since the quark-wall collision is soft. Correspondingly, the system chooses the QCD slingshot effect as the likely outcome.
    
Our results have a number of implications.  
First, as discussed, it allows us to capture certain important parallels between the behaviors of confined monopoles and quarks. In particular, in processes involving traversing the domain walls between confined and deconfined phases, both are expected to exhibit the slingshot effects.  
  
This effect can also have a number of important 
cosmological consequences since the above phases are expected to coexist at several stages of the universe's evolution. One observable imprint can occur in the form of gravitational waves.
Within this paper, we scrutinized the energy spectrum and emission direction of radiation from the slingshot scenario. Our observations reveal that the spectrum exhibits an $\omega^{-1}$ trend within our region of parameter space, akin to the behavior arising from the evolution of a cosmic string connecting a magnetic monopole and antimonopole~\cite{Martin:1996cp, Dvali:2022vwh}. Moreover, the emission takes place in a beaming angle in the direction of acceleration and scales, as a function of frequency as $\theta\propto \omega^{-1/2}$ within our range of parameters. 

In the last part of this work, we investigated the slingshot effect for the case of a vortex in $2+1$ dimensions, which can be extended to a theory with a cosmic string in $3+1$ dimensions that are confined by domain walls. 

Considering that cosmic strings are objects that can occur during a phase transition in the early universe, this scenario may also leave relevant marks in the gravitational wave background. Further explorations into this direction are left for future studies.\\
\newpage

    \section*{Acknowledgements}
This work was supported in part by the Humboldt Foundation under Humboldt Professorship Award, 
by the European Research Council Gravities Horizon Grant AO number: 850 173-6,  by the Deutsche Forschungsgemeinschaft (DFG, German Research Foundation) under Germany's Excellence Strategy - EXC-2111 - 390814868, and Germany's Excellence Strategy under Excellence Cluster Origins.\\[0.2cm]
\noindent\textbf{Disclaimer:}
Funded by the European Union. Views and opinions expressed are however those of the authors only and do not necessarily reflect those of the European Union or European Research Council. Neither the European Union nor the granting authority can be held responsible for them.

\setlength{\bibsep}{4pt}
\bibliography{references}

\end{document}